\documentclass[11pt]{iopart}

\usepackage[dvips]{graphicx}
\usepackage{psfrag}
\usepackage{epsfig}

\usepackage{bm}
\usepackage{bbm}

\renewcommand{\epsilon}{\varepsilon}

\newcommand{\ket}[1]{\! | #1\ \!\! \rangle}

\newcommand{\opbraket}[3]{ \langle #1    | #2 | #3  \rangle}
\newcommand{\gfbraket}[1]{\langle\!\langle #1  \rangle\!\rangle}

\newcommand{\unitop}{\mathbbm{1}}

\newcommand{\mod}[1]{\left | #1 \right |}

\newcommand{\expval}[2]{ \langle  #1 #2\ \!\! \rangle}

\newcommand{\elcre}[2]{ c^{\dagger}_{#1,#2}}
\newcommand{\elann}[2]{ c^{}_{#1,#2}}

\newcommand{\vct}[1]{\bm #1}
\newcommand{\vk}{{\bm k}}

\newcommand{\Imag}{\mathrm{Im}}

\newcommand{\hc}{\mathrm{h.c.}}

\begin{document}
\title{Spectral properties of locally correlated electrons in a BCS superconductor}


\author{J Bauer$^1$, A Oguri$^2$  and A C Hewson$^1$}
\address{$^1$Department of Mathematics, Imperial College, London SW7 2AZ, UK}
 \address{$^2$Department of Material Science, Osaka City University, Sumiyoshi-ku,
Osaka 558-8585 Japan}

\ead{j.bauer@imperial.ac.uk}

\date{\today} 

\begin{abstract}
We present a detailed  study of the spectral properties of a locally correlated
site embedded in a BCS superconducting medium. To this end the
Anderson impurity model with superconducting bath is analysed by
numerical renormalisation group (NRG) calculations. 
We calculate one and two-particle dynamic response function to
elucidate the spectral excitation and the nature of the ground state
for different parameter regimes with and without particle-hole symmetry.
The position and weight of the Andreev bound states  is
given for all relevant parameters. We also present phase
diagrams for the different ground state parameter regimes. 
This work is also relevant for dynamical mean field theory extensions
with superconducting symmetry breaking.


\end{abstract}

\section{Introduction} 

As described by Bardeen, Cooper and Schrieffer (BCS) \cite{BCS57} 
electrons in condensed matter with an attractive interaction assume a
superconducting state below a critical temperature, referred   
to as BCS state. In this state electrons with antiparallel spins   
form singlet bound states ($S=0$) known as Cooper pairs. This
pair formation is a fermionic many-body phenomenon as it relies on the
existence of a Fermi surface \cite{Coo56}. A singlet ground state
due to  many-body 
effects also occurs in a quite different situation,
when a magnetic impurity is embedded in a metallic host
\cite{Kon64,hewson}. This state, known as a Kondo singlet, occurs because  the electrons in the metal at low
temperature experience a large effective coupling to the localised impurity
spin. As a consequence it is energetically favourable to screen the local
moment, resulting in a (Kondo) singlet state ($S=0$).  

The BCS superconductivity and the Kondo effect,
are important topics in their own right, and have been extensively studied
by the condensed matter physics community.
The  interplay and competition of these two effects have also attracted a lot
of interest
because  metals with magnetic
impurities can   be superconducting at low temperatures
\cite{AG61,ZM70,Zit70,MZ71,Shi73,Mat77}.  The problem of dealing
with the two effects together is  complicated because
the magnetic impurities have a disruptive effect on the BCS
superconducting state and the Kondo
singlet formation leads to  a  breaking of the Cooper pairs. For a 
 recent review on this topic we refer to  \cite{BVZ06} and references
therein. Here we  address a particular aspect of the problem 
which has not so far received much attention, 
 the effects of the superconductivity on the local spectral properties of the 
 impurity. As in earlier studies, we take the
BCS superconductor as a fixed reference system and take as a model for the
impurity an interacting Anderson model. We employ the numerical
renormalisation group method (NRG), which is a reliable approach to
calculate low temperature spectral functions.

From earlier studies of this model, we know that if the interaction 
$U$ at the impurity site is weak, the ground state is dominated by the
superconducting behaviour and the singlet is predominantly 
a superconducting one. However, if there is a strong repulsion at the
impurity site, such that single occupation is
favoured, we have a situation where a single spin is coupled to the
superconducting medium. If the superconducting gap  $\Delta_{\rm
  sc}$ is very small then, similar  
  to the case with a normal, metallic bath, 
the ground state is a singlet, more specifically a Kondo
singlet. If this gap is increased, however, it is not possible
to form a Kondo singlet, due to the lack of states in the  vicinity of the Fermi
level,   and   the ground state becomes a doublet ($S=1/2$),
corresponding to an unscreened spin at the impurity site.
 This ground state
transition at zero temperature is an example of a quantum phase
transition which occurs for a level crossing that depends on a system
parameter \cite{sachdev}. The relevant energy scales for this
singlet-doublet transition to occur in the Kondo regime are the Kondo
temperature $T_{\rm K}$ and the superconducting gap $\Delta_{\rm
  sc}$. There have been numerical renormalisation group (NRG) studies
for the Kondo model \cite{SSSS92,SSSS93} and Anderson model
\cite{YO00} with superconducting bath. In these works the estimate for
the ground state transition is given by $T_{\rm K}/\Delta_{\rm sc}\simeq
0.3$, i.e. for $T_{\rm K}/\Delta_{\rm sc}> 0.3$ we have a singlet
ground state ($S=0$) whilst for $T_{\rm   K}/\Delta_{\rm sc}< 0.3$ the
ground state is a doublet. We can also consider the transition
 for a fixed
value of   $\Delta_{\rm
  sc}$ and increasing values of
the local interaction $U$.  In this case, as $U$ increases in the local moment regime,
$T_{\rm K}$ decreases until the singlet to doublet transition occurs
at a critical value $U=U_{c}$.

Due to the proximity effect there is an induced symmetry
breaking on the impurity site. As a consequence localised excited states (LES)
inside the superconducting gap can be induced at the impurity
site. Such states are well known from superconductor-normal-superconductor
(SNS) junctions and are usually called Andreev bound states. For a weak
on-site interaction the ground state of the system is usually a
superconducting singlet ($S=0$) and the LES is an $S=1/2$ excitation.
It is found that at the ground state transition the bound state energy
of the LES becomes zero as measured from the centre of the gap. This
is related to the fact that the level crossing occurs there.

In recent years detailed measurements on quantum dot structure have 
enabled one  to probe strong correlation effects
\cite{GSMAMK98,COK98}. In these experiments a quantum dot is coupled
to two leads, which can be superconducting.  In such situations finite
voltage induced currents \cite{RBT95,BNS02,BBNBBS03,VBMSY04} and  
Josephson currents \cite{DNBFK06}, induced by a phase difference,
were observed experimentally. For a theoretical description of this situation it is
important to characterise the Andreev bound states in the gap accurately.
Many of the more recent theoretical work \cite{RA99,Mat01,VMY03,SE04,CLKB04,OTH04,TOH07},  
focus on a quantum dot embedded in two superconducting baths with
different (complex) superconducting order parameters. 
These situations with two channels and with Josephson or nonequilibrium currents
will, however, not be covered in this paper. 

For the analysis presented here, which focuses on the spectral properties of
locally correlated electrons in the superconducting bath,
we use the NRG approach. We start by
outlining some of the details of the NRG calculation with a
superconducting medium in section 2. We also describe an analysis of
the Andreev bound states in the gap in terms of renormalised parameters, 
and discuss the limit of a large gap. In section 3 we present
results first for the model with particle-hole symmetry.
For low energies within the superconducting gap we calculate the
position and weight of the LES and also give the values for the
induced anomalous on-site correlation. We also present singlet-doublet
ground state phase diagrams for the symmetric and non-symmetric cases. 
The study is based on numerical renormalisation group (NRG)
calculations, which are capable of describing the full parameter range
from weak to strong coupling 
reliably. There have been a number of NRG studies of this situation in
the past \cite{SSSS92,SSSS93,YO00,CLKB04}. However, the dynamic
response function have not been addressed in a satisfactory way. Here we
present a thorough study of ground state and spectral properties,
which will also be of interest for cases where the AIM is used as an
effective model for superconductivity in the dynamical mean field
theory (DMFT) framework.

\section{The Anderson model with superconducting  medium} 
\label{sec:aimsc}
In the following we consider the Anderson impurity model (AIM) in the form 
\begin{equation}
  H=H_{d}+H_{\rm mix}+H_{\rm sc}.
\label{aimsc}
\end{equation}
The local part $H_{d}$, which describes an impurity or quantum dot, is given as usual by
\begin{equation}
H_{ d}= \sum_{{\sigma}}(\epsilon_{d}+\frac12U) \elcre{d}{\sigma}
 \elann{d}{\sigma} 
 + \frac12U\left(\sum_{\sigma}\elcre{d}{{\sigma}}
 \elann{d}{{\sigma}}-1\right)^2
\end{equation}
with the impurity level $\epsilon_{d}$ and an on-site interaction with
strength $U$. Also the mixing term has the usual form, 
\begin{eqnarray}
H_{\rm mix}
=\sum_{\vk,{\sigma}}V(\elcre{\vk}{\sigma}\elann{d}{\sigma} + \hc).
\end{eqnarray}
We define $\Gamma=\pi V^2\rho_c$ as the energy scale
for hybridisation, where $\rho_c=1/2D$ is the constant band
density of states of a flat band without superconducting symmetry breaking. The
superconducting medium is given in a BCS mean field form    
\begin{eqnarray}
H_{\rm sc}=\sum_{\vk,{\sigma}}\epsilon_{\vk} \elcre{\vk}{\sigma}
\elann{\vk}{\sigma} -  
\Delta_{\rm sc}\sum_{\vk}[\elcre{\vk}{\uparrow} \elcre{-\vk}{\downarrow}
+\hc],
\label{hscham}
\end{eqnarray}
where $\Delta_{\rm sc}$ is the isotropic superconducting gap
parameter, which is taken to be real for simplicity. In equation
(\ref{hscham}) the summations runs over 
all $\vk$ in a wide band. Another energy scale $\omega_{\rm D}$, the
Debye cutoff in BCS theory, could enter at this stage to restrict the
summation. As shown in reference \cite{SSSS92} with a scaling
argument, this effect does not alter the results substantially and
merely leads to slightly different parameters. The choice here
corresponds to $\omega_{\rm D}=D$, which was also assumed in  earlier
work \cite{SSSS92,YO00}. In appendix A we derive the equation for the
non-interacting local $d$-site Green's function matrix of the system 
(\ref{freegfctscimp}).

\subsection{The numerical renormalisation group (NRG) approach}
For the NRG approach we have to derive a discrete form of the
Hamiltonian, which can be 
diagonalised conveniently in a renormalisation group scheme descending to
lower energies. This is done in an analogous fashion as for a metallic medium
described in \cite{Wil75,KWW80a}. Essentially, there are
three steps which only 
affect $H_{\rm mix}$ and $H_{\rm sc}$: \\
(1) Mapping to a one-dimensional problem, (2) logarithmic
discretisation and (3) basis transformation. 
We obtain 
\begin{eqnarray}
H_{\rm mix}/D=
\sqrt{\frac{2\Gamma}{\pi D}}\sum_{{\sigma}}
(f^{\dagger}_{0{\sigma}}\elann{d}{\sigma}+  \hc), 
\end{eqnarray}
and 
\begin{eqnarray}
H^N_{\rm sc}/D=\sum^{N}_{{\sigma},n=0}\gamma_{n+1} 
(f^{\dagger}_{n{\sigma}}f_{n+1,{\sigma}} +  \hc)
-\frac{\Delta_{\rm
    sc}}{D}\sum^{N}_{n=0}(f^{\dagger}_{n\uparrow}f^{\dagger}_{n,\downarrow} 
+  \hc)  
\label{hscnrg}
\end{eqnarray}
where the parameters $\gamma_n$ have the usual form \cite{hewson}. For
more details we refer to  earlier work \cite{SSSS92,YO00}.
 
The iterative diagonalisation scheme is set up in the same way as in the
standard NRG case. Due to the anomalous term in the superconducting bath
$H^N_{\rm sc}$ the charge $Q$ is not a good quantum number of the
system. Thus eigenstates can only be characterised in terms of
the spin quantum number $S$. The coefficients $\gamma_n$ fall off with
$n$, but the second term in (\ref{hscnrg}) does not. Thus the
superconducting gap becomes a dominating energy scale for 
large $n$ and a relevant perturbation. It does not make sense to continue NRG
iterations down to energies much below this scale as there are no continuum
states anymore in the gap. Therefore, we stop the NRG procedure
at an iteration $N=N_{\rm max}$, such that the typical energy scale $\Lambda^{-(N_{\rm
    max}-1)/2}$ is not much smaller than the superconducting gap
$\Delta_{\rm sc}$. In practice the number of NRG iterations $N$ is
between 20-50 depending on the magnitude of the gap, where we chose
$\Lambda=1.8$ in all cases. We usually keep 800 states and the
$A_{\Lambda}$ factor \cite{KWW80a} is taken into account in the calculations. 

The NRG approach constitutes a reliable non-perturbative scheme to
calculate $T=0$ ground state properties of a local interacting many-body
problem. By putting together information obtained from different
iterations  dynamic response functions can also be obtained
\cite{hewson}. Here we calculate these spectral functions in the
approach \cite{PPA06,WD06pre} based on the complete Anders Schiller
basis \cite{AS05}. 
The Green's function of the  interacting
problem is given by the Dyson equation (\ref{scgreenfct}), which
involves the self-energy matrix $\underline \Sigma(\omega)$.
In appendix B we describe how the diagonal part of the self-energy 
$\Sigma(\omega)=\Sigma_{11}(\omega)$ and the offdiagonal part of the self-energy
$\Sigma^{\rm off}(\omega)=\Sigma_{21}(\omega)$ can be calculated from
dynamic response functions in the NRG calculation, which is in analogy to
the method described in reference \cite{BHP98}.

\subsection{The Andreev bound states}
\label{sec:andbs}
The denominator of the $d$-site Green's function, equation
(\ref{scgreenfct}), can vanish inside the gap
$\mod{\omega}<\Delta_{\rm sc}$.  
As the imaginary part of the self-energy is zero in the gap this leads to 
excitations with infinite lifetime there. They correspond to the localised
excited states (LES) or Andreev bound states. For the non-interacting case
they are determined by the equation $D(\omega)=0$ [cf. eq. (\ref{scdet})],
\begin{equation}
  \label{noniaandbs}
  \omega^2-\epsilon_d^2-\Gamma^2+\frac{2\omega^2\Gamma}{E(\omega)}=0,
\end{equation}
where the function $E(\omega)$ is given in equation (\ref{funcEom}).
The terms in equation (\ref{noniaandbs}) are functions of $\omega^2$, so if $E^0_b$ is a
solution so is $-E^0_b$.
In general, in the interacting case we have to analyse the equation
\begin{equation}
\fl
\Big[\omega-\epsilon_d+\frac{\omega\Gamma}{E(\omega)}-\Sigma(\omega)\Big]
\Big[\omega+\epsilon_d+\frac{\omega\Gamma}{E(\omega)}+\Sigma(-\omega)^*\Big]
-\Big[\frac{\Gamma\Delta_{\rm sc}}{E(\omega)}-\Sigma^{\rm off}(\omega)\Big]
\Big[\frac{\Gamma\Delta_{\rm sc}}{E(\omega)}-\Sigma^{\rm
  off}(-\omega)^*\Big]=0. 
\label{intactBE}
\end{equation}
Once the self-energies are calculated it is possible to solve this equation
iteratively. Here, we will develop a simplified description by using a low
energy expansion of the self-energy. First note that in the gap,
$\mod{\omega}<\Delta_{\rm sc}$, $\Imag\Sigma(\omega)=\Imag\Sigma^{\rm
  off}(\omega)=0$. We expand the real part of the diagonal self-energy 
$\Sigma(\omega)$ to first order around $\omega=0$, which is motivated by the
Fermi liquid expansions for the normal metallic case and justified by
the numerical results for the behaviour for low frequency. The
offdiagonal self-energy is approximated simply by the real constant
$\Sigma^{\rm off}(0)$. This approximation for the self-energy is easy
to justify if  the gap is small parameter, such that it only covers
small values of $\omega$. The main objective is to present a
simplified picture for the analysis of the Andreev bound state for the
interacting system. We do not expect to be able to describe the
system near the quantum phase transition accurately like this, and
other limitations will be seen in the results later. 
Hence, we find instead of (\ref{intactBE}) the simpler equation   
\begin{equation}
\label{iaandbs}
  \omega^2-\tilde\epsilon_d^2-\tilde\Gamma^2-z^2\Sigma^{\rm off}(0)^2 
 +\frac{2\tilde\Gamma[\omega^2+\Delta_{\rm sc}z\Sigma^{\rm off}(0)]}{E(\omega)}=0 ,
\end{equation}
where renormalised parameters $\tilde\epsilon_d=z[\epsilon_d+\Sigma(0)]$ and
$\tilde \Gamma=z\Gamma$ were introduced.  As usual $z^{-1}=1-\Sigma'(0)$.
Renormalised parameters for
the analysis of the Andreev bound states were also considered in
reference \cite{VMY03,YMV03}. The definition here corresponds to the
renormalised perturbation theory framework for the AIM introduced in
\cite{Hew93}.
The form of the equations (\ref{noniaandbs}) and (\ref{iaandbs}) is very
similar and both can be easily solved numerically to give the bound state
solutions $\omega=E^{\alpha}_b=\alpha E_b$, $\alpha=\pm$. Due to the
additional offdiagonal correlations induced by the self-energy term $\Sigma^{\rm
  off}(0)$,  a simple interpretation of the interacting theory 
based on using renormalised  parameters $\tilde\epsilon_d$, $\tilde \Gamma$ in
equation (\ref{noniaandbs}) for the non-interacting theory is, however, not possible.

Based on the same idea we can give approximate expressions for the weights of
the bound states $w_b^{\alpha}$ by expanding the diagonal part of the Green's
function around $\omega=E^{\alpha}_b$. We can write the retarded Green's function in
the gap near the bound states $\omega\simeq\pm E_b$ as  
\begin{equation}
  G(\omega)=\frac{w^{-}_b}{\omega-E_b^-+i\eta}+\frac{w^{+}_b}{\omega-E_b^++i\eta}.
\end{equation}
Using the above approximation for the self-energy the weights are found to be
\begin{equation}
  \label{weightsbs}
  w_b^{\alpha}=\frac{z}2
  E(E_b)^2\frac{E(E_b)(1+\alpha\frac{\tilde\epsilon_d}{E_b})+\tilde\Gamma}
 {E(E_b)^2(E(E_b)+2\tilde\Gamma)+\tilde\Gamma(E_b^2+\Delta_{\rm
  sc}z\Sigma^{\rm off}(0))}.\label{wts}
\end{equation}
In a more sophisticated approximation one could consider an expansion of the
self-energies around the bound state energies $E_b$ rather than $\omega=0$.
Various things can be inferred from expression (\ref{weightsbs}). First we
note that in the particle-hole symmetric case, $\tilde \epsilon_d=0$,
$w_b^{+}=w_b^{-}=w_b$. The weights are proportional to the
renormalisation factor $z$. Since $z$ shows a similar behaviour as in
the metallic lead case they decrease with increasing interaction
$U$ according to (\ref{weightsbs}). One can easily see that for 
bound state energies close to the gap, $\mod{E_b}\to\Delta_{\rm sc}$, the weights go
to zero, $w_b^{\alpha}\to 0$. 
One finds \cite{YO00} that for small $U/\pi\Gamma$ and $\Delta_{\rm
  sc}/\Gamma\ll 1$ we have $E_b\to\Delta_{\rm sc}$, and also for large
$U/\pi\Gamma$ the bound state energy is close to the gap. Therefore the
overall behaviour for $w_b$ is given in such a case by $w_b\to0$ for
small $U$, then an increase with $U$ to a  maximum and a decay again
for large $U$ [cf. figure \ref{boundstates} later]. 
At the ground state transition, where $E_b=0$, the weight shows a
discontinuity, and from equation (\ref{weightsbs}) this requires a
jump of the self-energy as function of $U$.

\subsection{The limit of large gap}
\label{sec:limlargegap}
In order to obtain some obtain analytical results it is useful is to
consider the case where the superconducting gap is a large parameter,
$\Delta_{\rm sc}\to \infty$ 
\cite{VMY03,OTH04,TOH07,ACZ00}. Then the problem essentially reduces to a localised
model with an anomalous on-site term which is of the order of the
hybridisation $\Gamma$. We will write it in the form  
\begin{equation}
  H^{\rm \infty}_{\rm d}=\sum_{{\sigma}}\xi_d (\elcre{d}{\sigma}\elann{d}{\sigma}-1) -
  \Gamma[\elcre{d}{\uparrow} \elcre{d}{\downarrow} +\hc]+\frac
  U2\Big(\sum_{\sigma}n_{d,\sigma}-1\Big)^2, 
\label{scsingsiteham}
\end{equation}
where $\xi_d=\epsilon_d+U/2$. Without interaction this Hamiltonian can be
diagonalised by a Bogoliubov transformation and the excitation energies
$E_d=\sqrt{\xi_d^2+\Gamma^2}$ are found, which lie in the gap as
$\Gamma\ll\Delta_{\rm sc}$ as assumed initially. This gives a direct picture of the
emergence of the Andreev bound states for large $\Delta_{\rm sc}$.

We can discuss the ground state crossover from the singlet to the doublet
state in terms of the single site Hamiltonian (\ref{scsingsiteham}). First
note that the $S=1/2$ (doublet) states, $\ket{\!\!\uparrow}$ and
$\ket{\!\!\downarrow}$, are eigenstates of (\ref{scsingsiteham}) with
zero energy. The $S=0$ singlet states, empty site $\ket{0}$ 
and doubly occupied site $\ket{\!\!\uparrow\downarrow}$, are not
eigenstates of (\ref{scsingsiteham}). However, the linear combinations
in  the ``BCS-form'',
\begin{equation}
  \ket{\Psi_1}=u_d\,\ket{0}+v_d\,\ket{\!\!\uparrow\downarrow},
\qquad
\ket{\Psi_2}=v_d\,\ket{0}-u_d\,\ket{\!\!\uparrow\downarrow},
\label{wffct}
\end{equation}
are eigenstates with eigenvalues $E_1=-E_d+U/2$ and $E_2=E_d+U/2$,
respectively. The coefficients $u_d,v_d$ are given by 
\begin{equation}
  u_d^2=\frac12\Big(1+\frac{\xi_d}{E_d}\Big),
\qquad
  v_d^2=\frac12\Big(1-\frac{\xi_d}{E_d}\Big).
\end{equation}
The ground-state is therefore a singlet as long as $E_1<0$ and a doublet
otherwise. The condition $E_1=0$ or
\begin{equation}
\frac{\xi_d^2 }{U^2}+\frac{\Gamma^2 }{U^2}=\frac14
\label{delinfphasbound}
\end{equation}
defines therefore the phase boundary for the transition. It is a semicircle in the
$(\xi_d/U)$-$(\Gamma/U)$-plane with radius $1/2$, which is shown in figure
\ref{phasediag} later. How this phase boundary looks like for finite gap
$\Delta_{\rm sc}$ will be investigated in section \ref{sec:scawphsym},
when we look at the situation away from particle-hole symmetry. 
In the case of particle-hole symmetry $\xi_d=0$ and condition
(\ref{delinfphasbound}) reduces to $\Gamma=U/2$. 

Having established the formalism and the most important relations we
will in the next section present results for spectral behaviour of the
symmetric AIM with superconducting bath with a finite gap parameter.

\section{Results}

In this section we present results for the local spectral
properties. The diagonal and offdiagonal Green's functions are
calculated within the NRG framework usually from the Lehmann
representation, 
\begin{equation}
  \rho_{d}(\omega)=\frac1Z\sum_{m,n}|\opbraket
  m{c_{d}^{\dagger}}n|^2\delta[\omega-(E_m-E_n)]
 (\e^{-\beta E_m}+\e^{-\beta E_n}),
\label{speclehman}
\end{equation}
and similar for the offdiagonal Green's function.
As in this procedure the discrete excitations for the spectral peaks in the
Green's functions have to be broadened, it is not straight forward like this to obtain the
  sharp spectral gap at $\mod{\omega}=\Delta_{\rm sc}$ expected for $T=0$ .  
As detailed in appendix B, we can, however, determine the self-energy
matrix from the one-particle Green's function and the higher $F$-Green's
function [cf. eq. (\ref{SigF})]. Then we can use the exact expression for the  
non-interacting Green's function $\underline {G}^0_d(\omega)$ in equation
(\ref{freegfctscimp}), which includes a sharp spectral gap, 
and the Dyson matrix equation (\ref{scgreenfct}) to calculate the diagonal and
offdiagonal Green's function, $G(\omega)$ and $G^{\rm off}(\omega)$
respectively. This is the way the Green's functions are calculated for
the region outside the gap, $\mod{\omega}>\Delta_{\rm sc}$. Inside the
gap, $\mod{\omega}<\Delta_{\rm sc}$, we have extracted the weights
$w_b$ and positions $E^{\alpha}_b$ of the delta-function peaks for the Andreev
bound states from the NRG excitation data for the Green's
function directly from the lowest spectral excitation (SE) in equation
(\ref{speclehman}).  
These delta-functions are represented by an arrow in the plots.
Altogether the diagonal spectral function $\rho(\omega)=-\Imag
G(\omega)/\pi$ can then be written in the form   
\begin{equation}
  \rho(\omega)=\sum_{\alpha=\pm}w_b\delta(\omega- E^{\alpha}_b)+\rho_{\rm
  cont}(\omega),
\label{diagspecfct}
\end{equation}
where $\rho_{\rm cont}(\omega)$ is the continuum part for $\mod{\omega}>\Delta_{\rm sc}$.
The offdiagonal part of the spectrum $\rho^{\rm off}(\omega)=-\Imag G^{\rm
  off}(\omega)/\pi$ has a similar general form as the diagonal part,
\begin{equation}
  \rho^{\rm off}(\omega)=\sum_{\alpha=\pm}\bar w^{\alpha}_b\delta(\omega-
  E^{\alpha}_b)+\rho^{\rm off}_{\rm 
  cont}(\omega),
\label{offdiagspecfct}
\end{equation}
where the weights $\bar w^{\alpha}_b$ can have positive and negative
values. For half filling the spectrum $\rho^{\rm off}(\omega)$ is an
asymmetric function of $\omega$.

\subsection{Symmetric model}

We first focus on the particle-hole symmetric model,
$\epsilon_d=-U/2$, where the ratio $U/\pi\Gamma$ and the parameter
$\Delta_{\rm sc}$ are the relevant energy scales.
\subsubsection{Spectral functions for small gap} 
\begin{figure}[!htbp]
\centering
\includegraphics[width=0.45\textwidth]{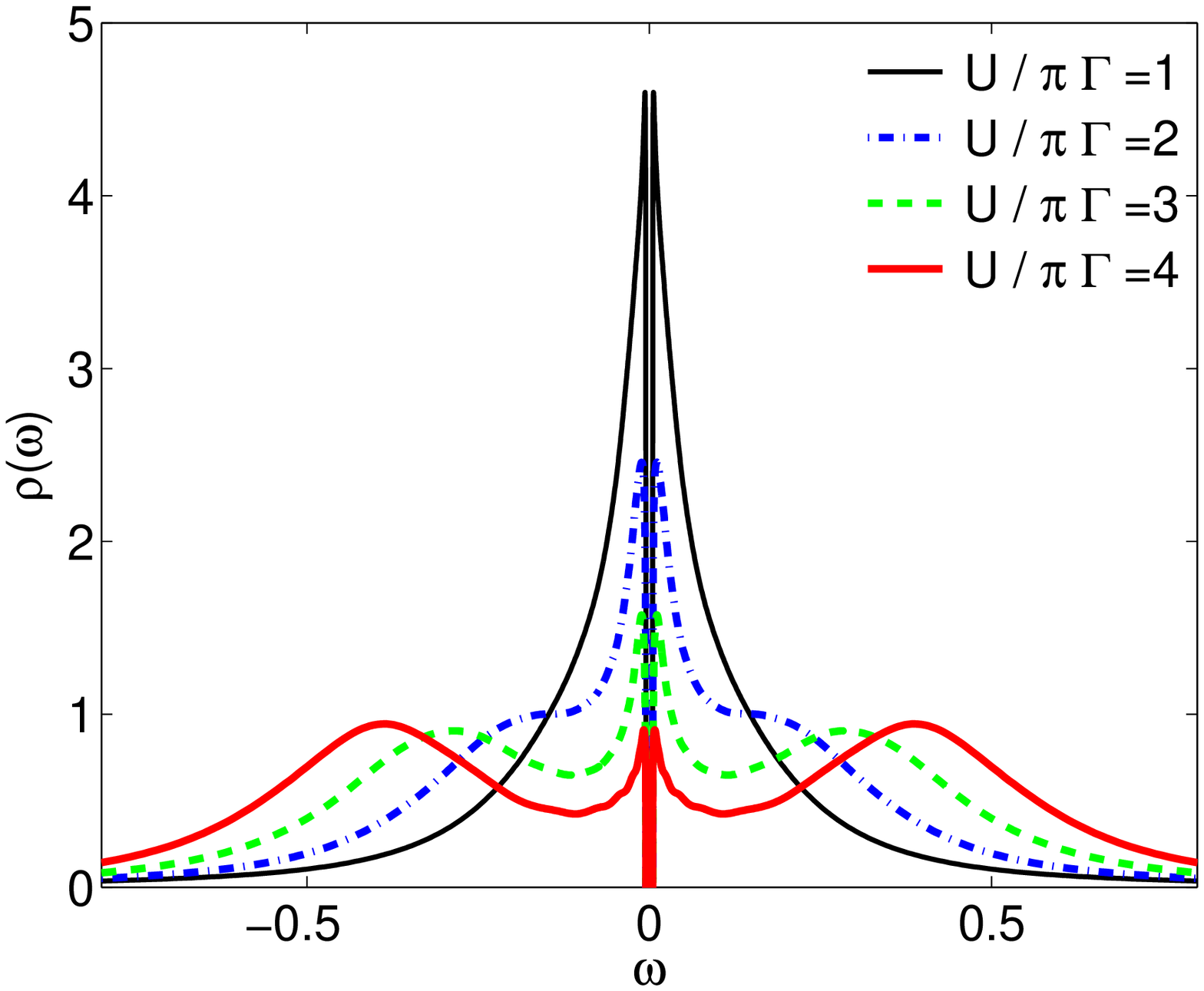}
\hspace{0.5cm}
\includegraphics[width=0.45\textwidth]{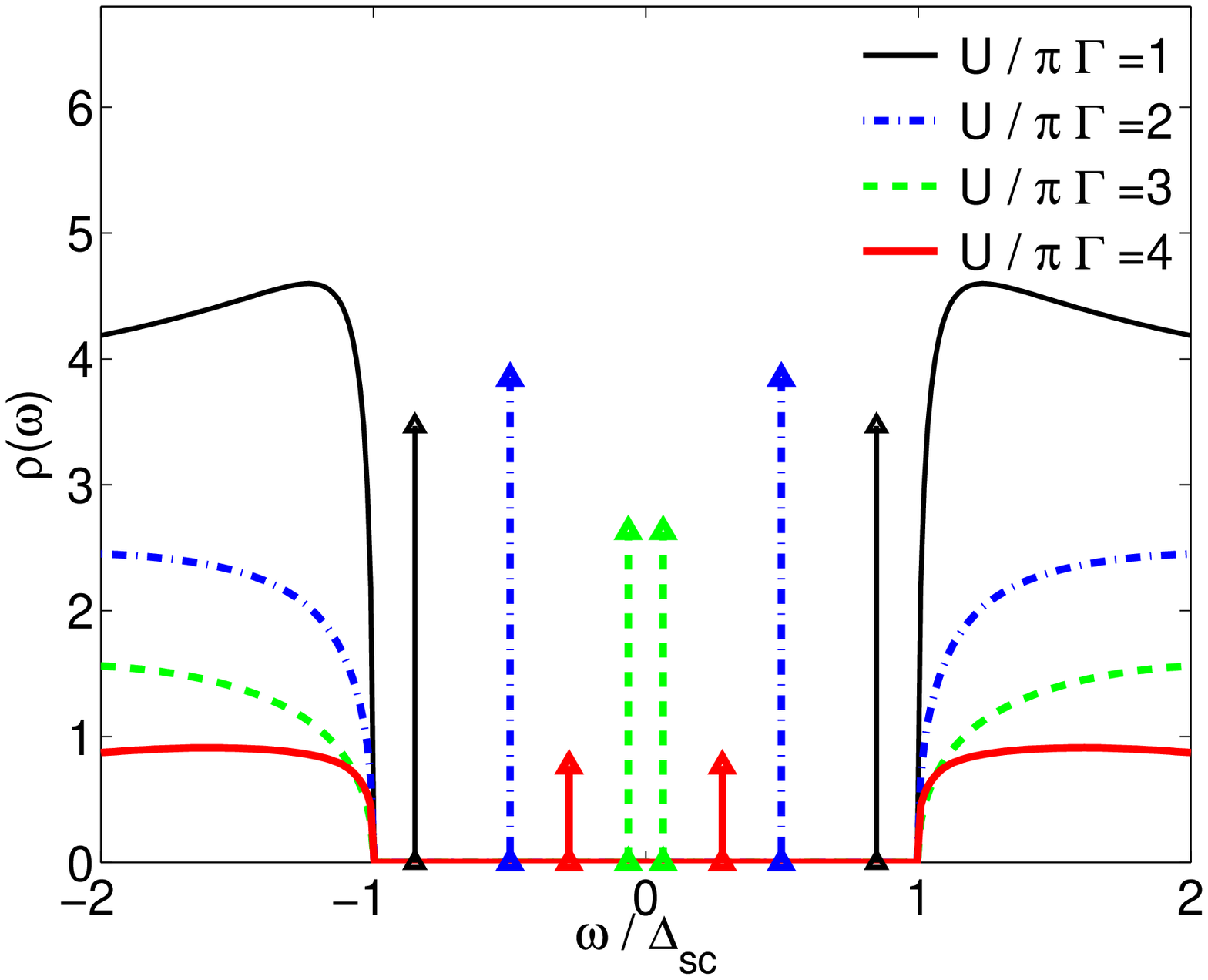}
\vspace*{-0.5cm}
\caption{The spectral density $\rho(\omega)$ for various values of $U$ for the
  whole energy regime (left) and the region in the gap (right);
  $\Delta_{\rm sc}=0.005$ and $\pi\Gamma= 0.2$.}    
\label{diagspecdel0.005}
\end{figure}

In figure \ref{diagspecdel0.005} we show the 
spectral function (\ref{diagspecfct}) for $\Delta_{\rm sc}=0.005$ for the
diagonal Green's function at the impurity site for a number of different
values of $U$. Here and in the following we take a fixed value for the hybridisation,
$\pi\Gamma=0.2$. All quantities can be thought of as being scaled by half the
band width $D=1$.   

\noindent 
In the plot on the left hand side we give the spectrum over the full energy
range. When the interaction is increased, spectral weight is shifted to
higher energies as the atomic limit peaks at $\pm U/2$ develop . We
also observe the beginning of the formation of a Kondo resonance at
low frequencies.  For larger $U$ the Kondo resonance becomes narrower, but
its formation is suppressed in the very low frequency regime
 because the spectral density vanishes in the gap
region $-\Delta_{\rm sc}<\omega<\Delta_{\rm sc}$. This is not visible on the
  scale used in the left hand panel of figure \ref{diagspecdel0.005}.
In the right hand panel of figure \ref{diagspecdel0.005} we give an 
enlarged plot of the gap region,
 which shows the delta function contributions from  the Andreev bound
states,
where the arrows give the position of the bound state $E^{\pm}_b$ and their
height indicates the spectral weight $w_b$. It can be seen that the
position of the bound state changes when we increase the interaction. The
weight first increases and then decreases as a function of $U$, which
corresponds to the feature which was interpreted earlier using
equation (\ref{wts}). It is generally of interest to see how much
spectral weight is transfered from the continuum to the bound states,
and an overview for this is given in the later figure \ref{ddexp}
(right) and \ref{hfphasedia}.
Note that the largest value of $U$ shown, is greater than the critical $U_c$
for the singlet-doublet transition ($U_c/\pi\Gamma\simeq 3.2$). In the high
energy spectrum there is no significant change to be seen in the behaviour,
however, at low energies we observe the crossing of the bound state energies at
$\omega=0$ at $U_c$.  

\begin{figure}[!htbp]
\centering
\includegraphics[width=0.45\textwidth]{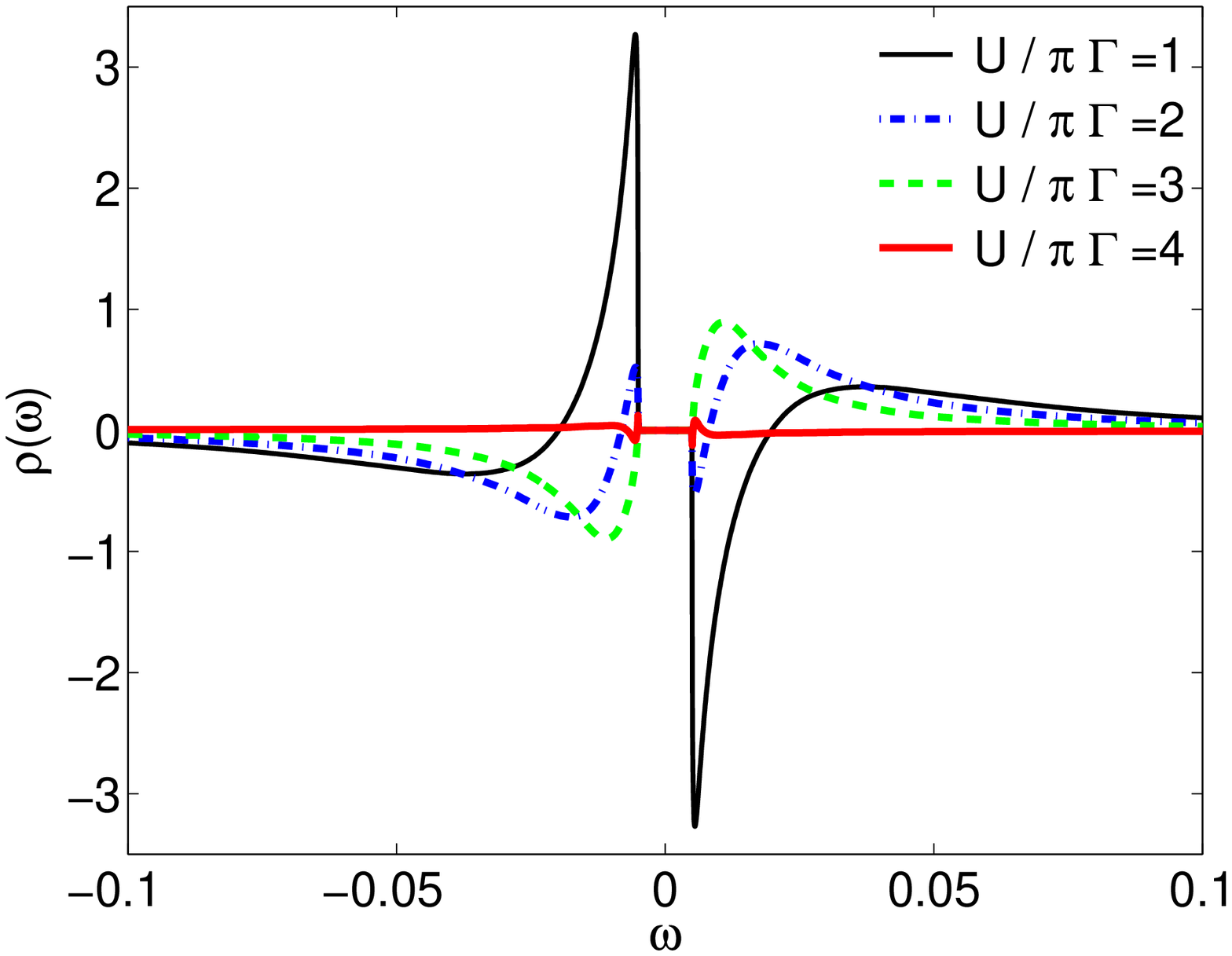}
\hspace{0.5cm}
\includegraphics[width=0.45\textwidth]{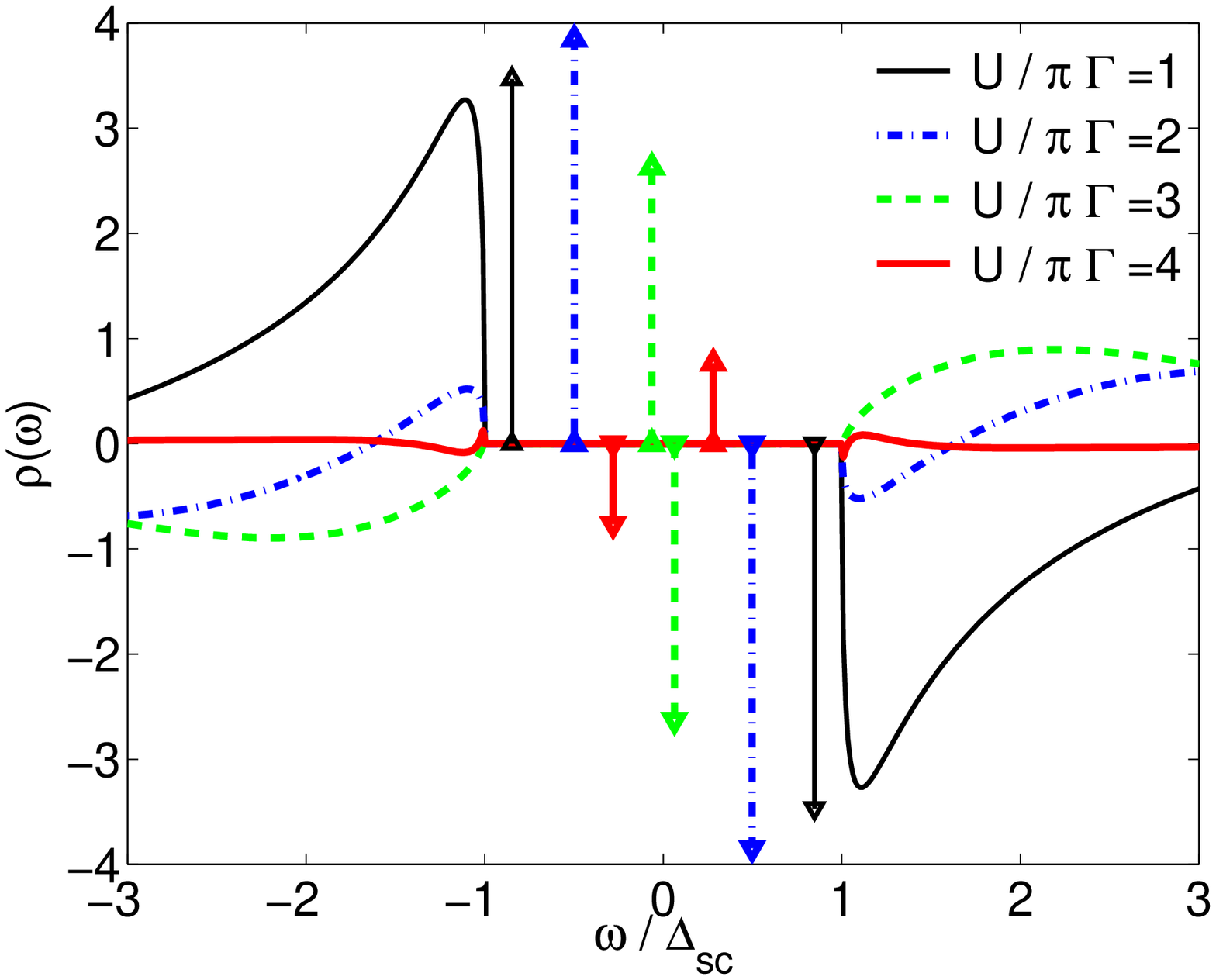}
\vspace*{-0.5cm}
\caption{The spectral density $\rho^{\rm off}(\omega)$ for various values of $U$ for the
  whole energy regime (left) and the region in the gap (right);
  $\Delta_{\rm sc}=0.005$ and  $\pi\Gamma= 0.2$.}    
\label{offdiagspecdel0.005}
\end{figure}
\noindent 
In figure \ref{offdiagspecdel0.005} we show
the offdiagonal spectral function (\ref{offdiagspecfct}) for $\Delta_{\rm sc}=0.005$
for a number of different values of $U$.   
In the plot on the left hand side we show the behaviour for the continuum
part outside the gap. Notice that the frequency range only extends up to
$\omega=\pm 0.1$. We can see a peak close to $\omega=\pm \Delta_{\rm sc}$, which is
suppressed for larger $U$ and changes sign towards the singlet-doublet transition.
The behaviour of the bound state peaks in the offdiagonal spectrum is
displayed on the right hand side of the figure. We can see similar features as
observed before in the diagonal part, i.e. the weight first increases with $U$ and then
decreases. If we follow the excitations with the weight of the same sign we
can see, that at the singlet-doublet transition the bound state levels cross
at $\omega=0$.

\subsubsection{Bound state behaviour} 
\begin{figure}[!htbp]
\centering
\includegraphics[width=0.45\textwidth]{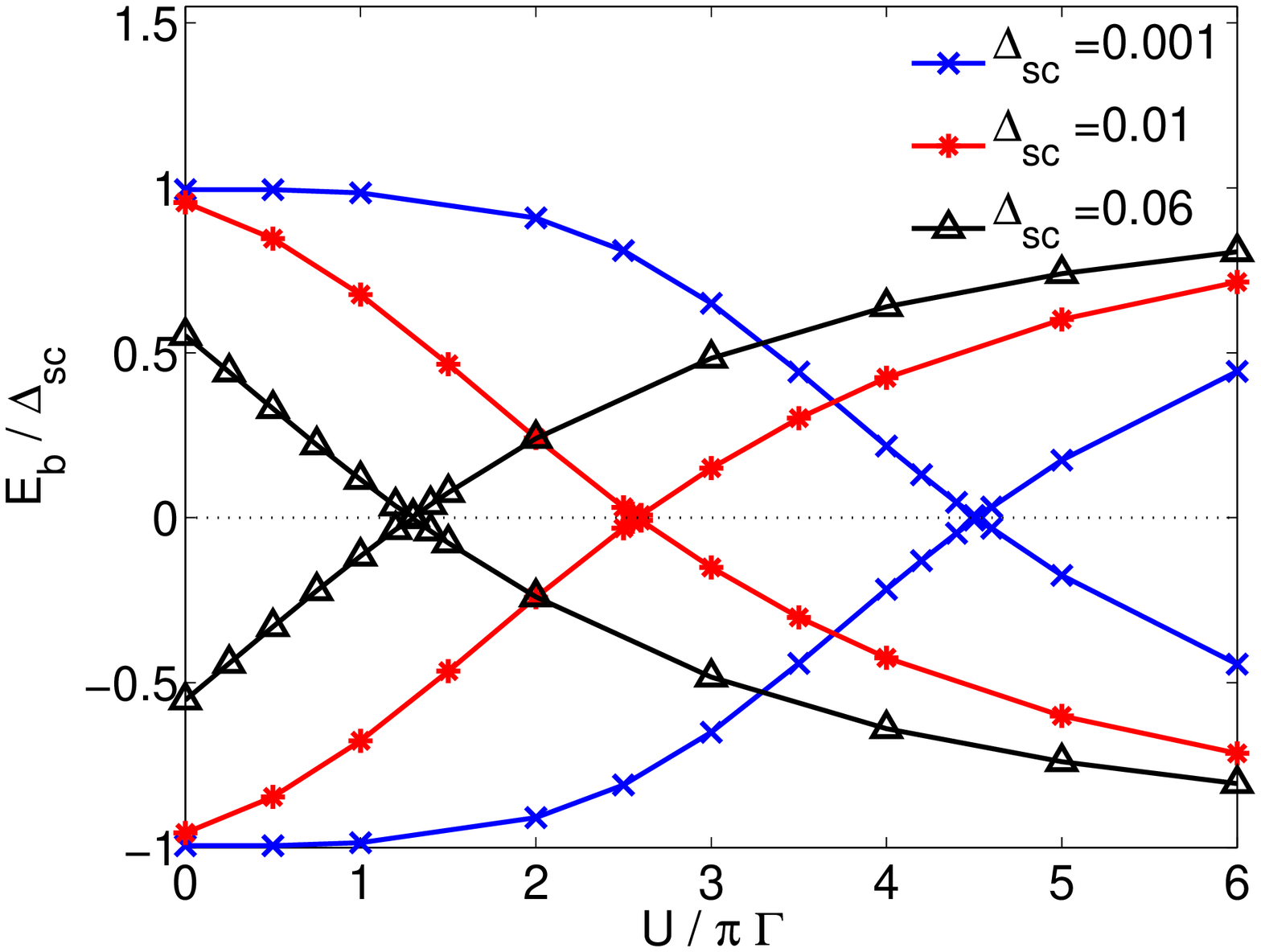}
\hspace{0.5cm}
\includegraphics[width=0.45\textwidth]{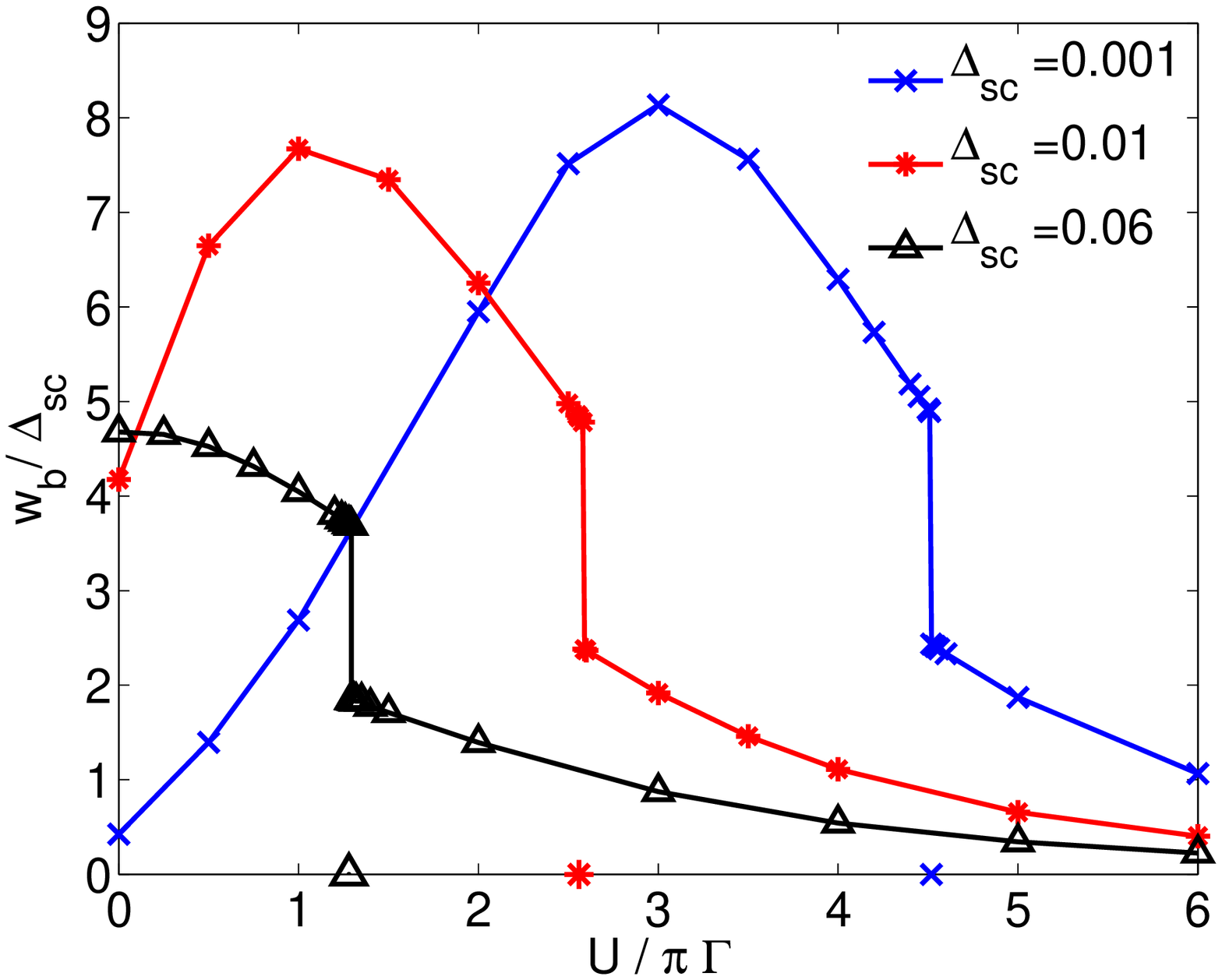}
\vspace*{-0.5cm}
\caption{Bound state energies $E_b$ (left) and weights $w_b$ (right) for various $U/\pi
  \Gamma$ and $\Delta_{\rm sc}$. Both quantities have been scaled by the
  corresponding value of $\Delta_{\rm sc}$;  $\pi\Gamma= 0.2$.}     
\label{boundstates}
\end{figure}

A more detailed analysis of the behaviour of the bound state 
as a function of $U/\pi\Gamma$ and the gap in the medium $\Delta_{\rm sc}$
is presented in figure \ref{boundstates}.
On the left hand side we plot the bound state energies $\pm E_b$ and
on the right hand side the corresponding weights $w_b$. 

\noindent
We can see that in the non-interacting case the bound state energy for the
cases with small gap ($\Delta_{\rm sc}=0.001,0.01$) is very close to 
$\pm\Delta_{\rm sc}$ and decreases to zero
with increasing interaction. For a critical value $U_c$ the nature of the
ground-state changes from a singlet ($S=0$) to a doublet ($S=1/2$) and at this
point $E_b=0$. For this transition we can think of the positive $E^+_b$ and negative
solution $E^-_b$ for the bound states as crossing at $\omega=0$. When 
the interaction is increased further, $\mod{E^{\pm}_b}$ becomes finite again and increases
with $U$. The larger the gap $\Delta_{\rm sc}$ the smaller critical value
$U_c$ for this ground state transition becomes. 
In the case where $\Delta_{\rm sc}$ is of the order of $\Gamma$ - as can be seen
for the case $\Delta_{\rm sc}=0.06$ -  the  
bound state energy $E_b$ lies in the middle of the gap already for the
non-interacting case, but otherwise shows a similar behaviour as described above.

On the right hand side of figure \ref{boundstates} the weight
$w_b$ of these bound states can be seen. We have marked the position $U_c$ of the
singlet-doublet crossover point by a symbol on the $x$-axis. The two curves
for a value of the gap $\Delta_{\rm sc}=0.001$ and $\Delta_{\rm sc}=0.01$ 
have a maximum for some intermediate value of $U$ which is smaller than the
critical $U_c$ for the ground state transition. This behaviour can be
understood from the analytic behaviour of the explicit equation
(\ref{weightsbs}) derived earlier. For the other curve ($\Delta_{\rm
  sc}=0.06$) the weight is maximal for the non-interacting case. In
all cases the weight becomes very  small for large $U$. Note that we
plot the weight scaled by the gap parameter, $w_b/\Delta_{\rm   sc}$, and
therefore the absolute values are larger for the cases with larger
superconducting gap. At the singlet-doublet transition we can see
discontinuous behaviour as the weight changes sharply. This is a
feature of the zero temperature calculation, where the matrix elements
in the Lehmann sum (\ref{speclehman}) change their values
discontinuously when the levels cross on increasing $U$,
such that the nature of the ground state changes. It can be seen for
the anomalous correlations $\langle d_{\uparrow}d_{\downarrow}\rangle$
in figure \ref{ddexp} later, as well. For finite temperature this
discontinuity  becomes smooth. 

\subsubsection{Spectral functions for larger gap} 

\begin{figure}[!htbp]
\centering
\includegraphics[width=0.45\textwidth]{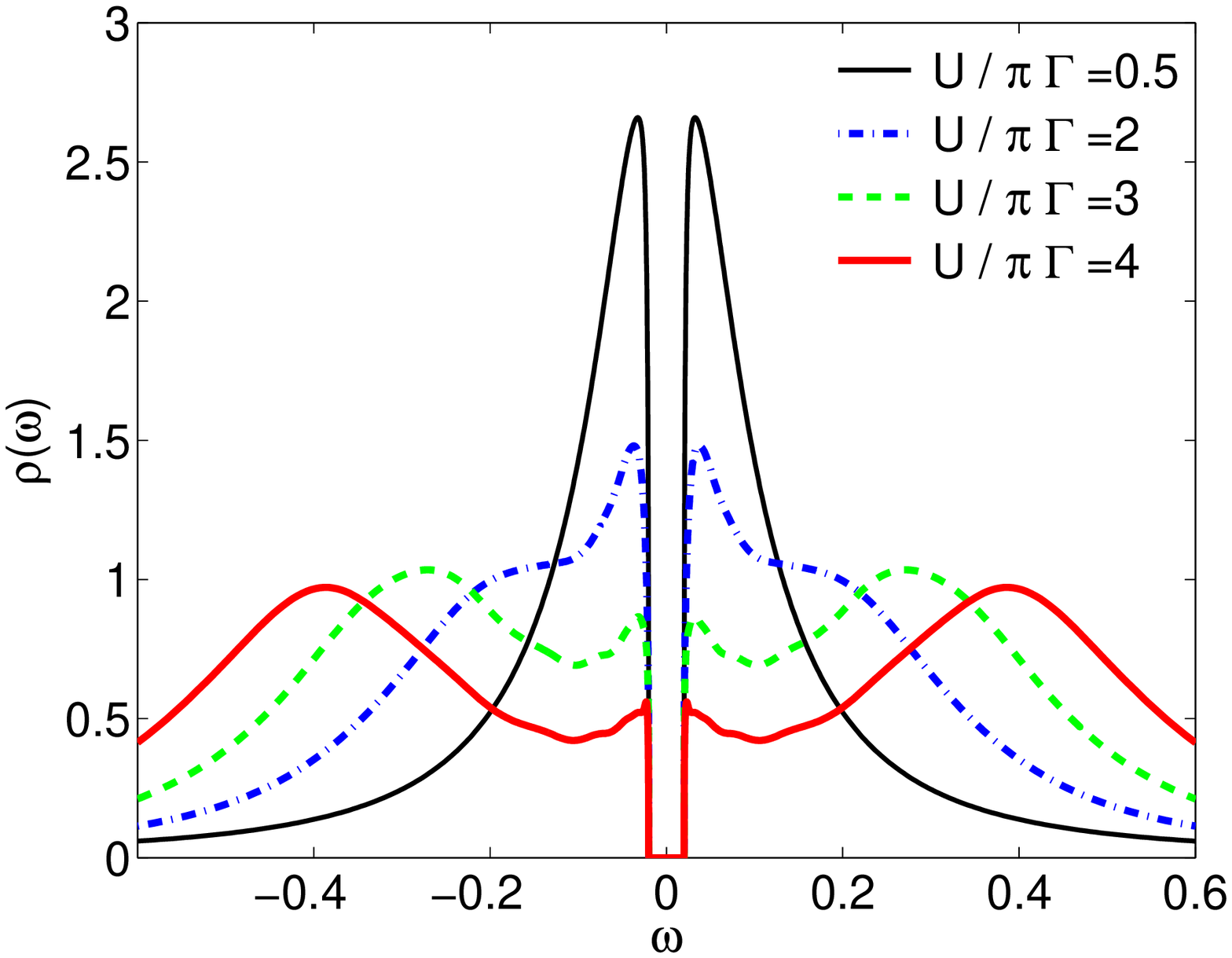}
\hspace{0.5cm}
\includegraphics[width=0.45\textwidth]{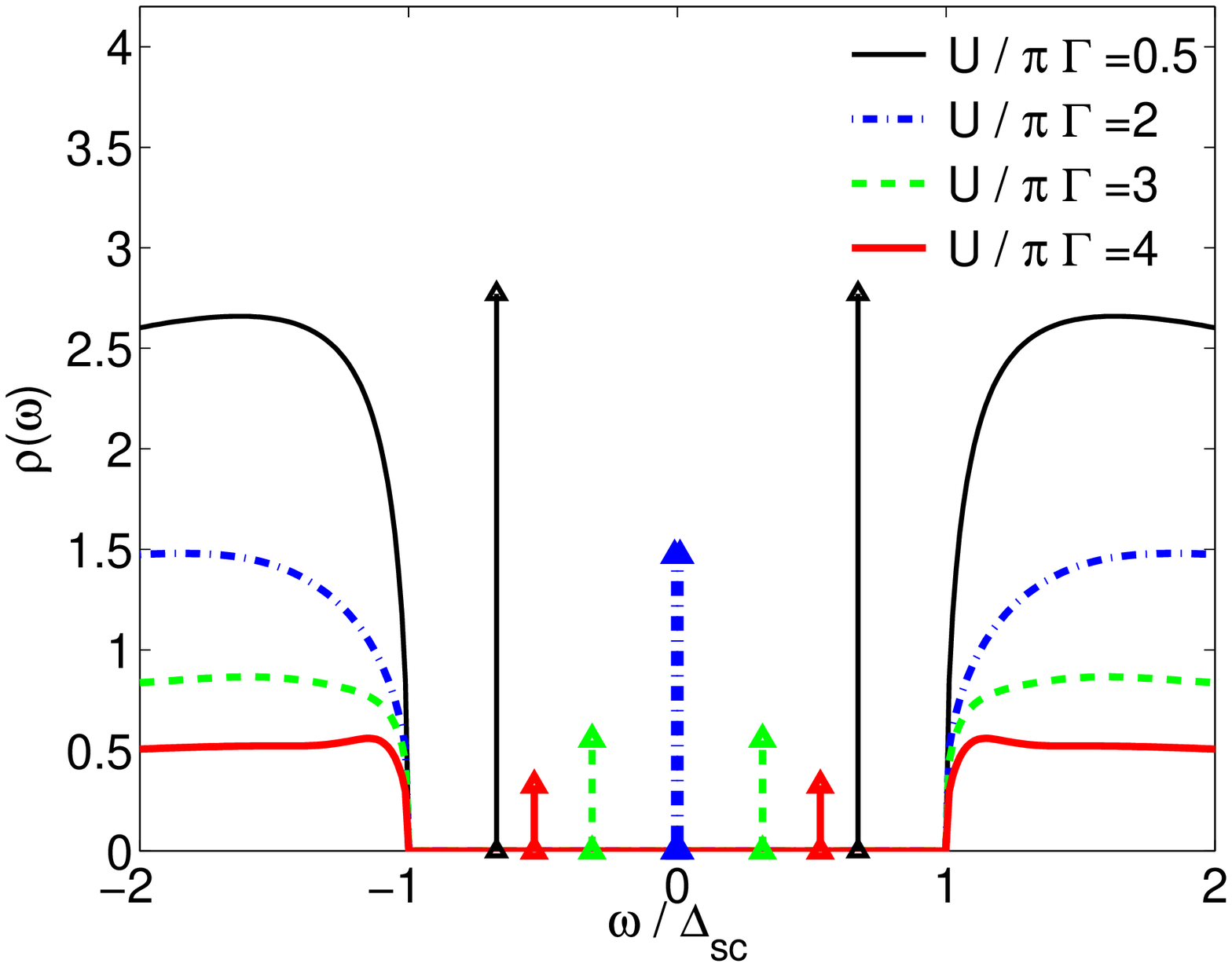}
\vspace*{-0.5cm}
\caption{The spectral density $\rho(\omega)$ for various values of $U$ for the
  whole energy regime (left) and the region in the gap (right);
  $\Delta_{\rm sc}=0.02$ and $\pi\Gamma= 0.2$.}    
\label{diagspecdel0.02}
\end{figure}

In figure \ref{diagspecdel0.02} we show for comparison the diagonal
spectral function for a larger gap $\Delta_{\rm sc}=0.02$ for the
diagonal Green's function at the impurity site for a number of
different values of $U$.   

\noindent
The overall picture on the left is similar to the case in figure
\ref{diagspecdel0.005} with the smaller gap. Due to the larger gap the
formation of the central Kondo resonance is completely suppressed, but the
high energy spectrum is as before. From the  behaviour within the gap (right
side in figure \ref{diagspecdel0.02}) we can see that the bound state position
$E^{\pm}_b$ goes to zero for smaller $U$ values than in the case $\Delta_{\rm
  sc}=0.005$, and hence the ground state transition occurs for smaller
$U_c$ for the larger gap ($U_c/\pi\Gamma\simeq 2.03$). This was
analysed in detail in figure \ref{boundstates} above. For the
values of $U$ shown the spectral weight of the bound states $w_b$ 
decreases with increasing $U$. The weight $w_b$ of the peaks in the
gap has been scaled differently in figures \ref{diagspecdel0.005} and
\ref{diagspecdel0.02}, so that their height should not be compared directly. 

\begin{figure}[!htbp]
\centering
\includegraphics[width=0.45\textwidth]{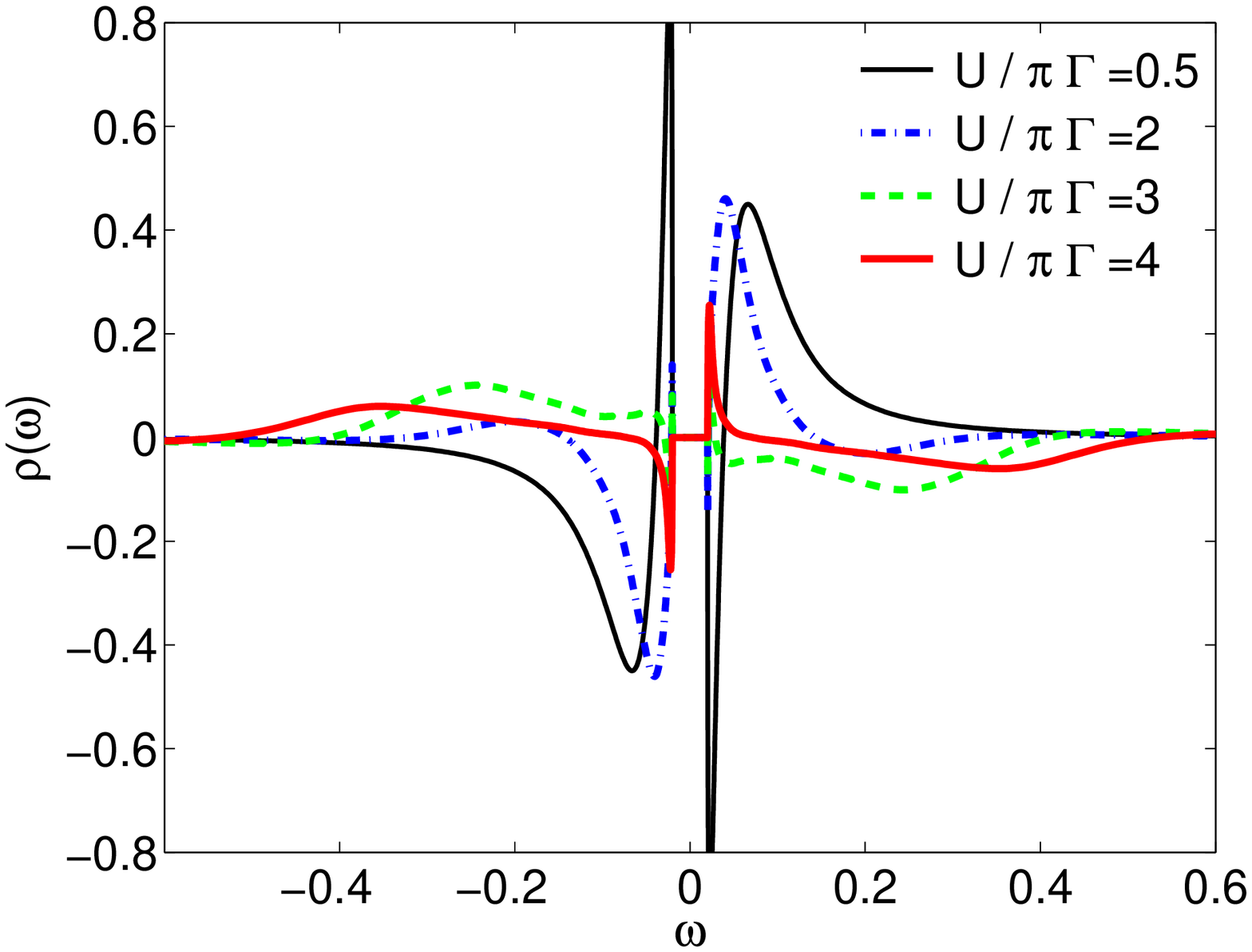}
\hspace{0.5cm}
\includegraphics[width=0.45\textwidth]{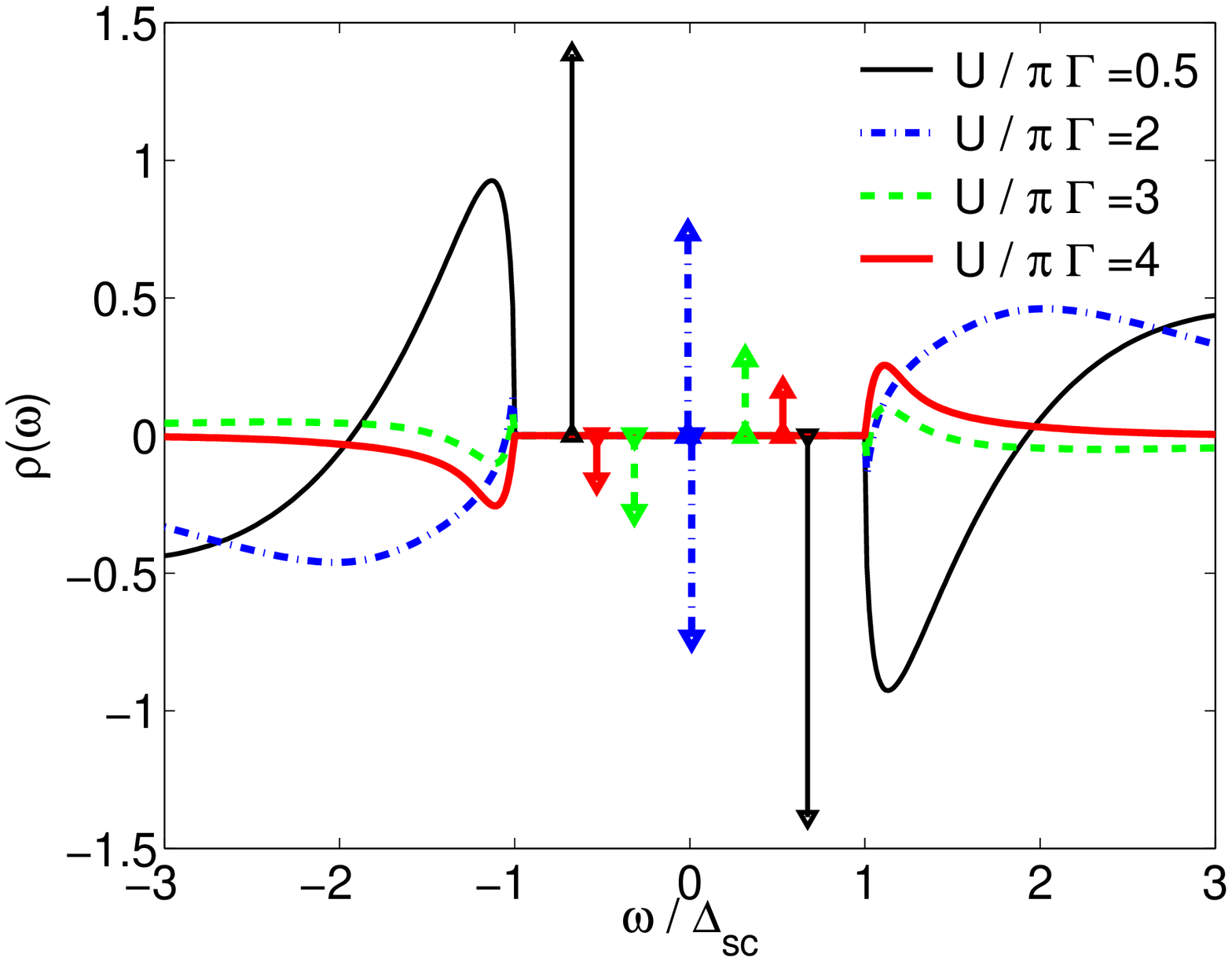}
\vspace*{-0.5cm}
\caption{The spectral density $\rho^{\rm off}(\omega)$ for various values of $U$ for the
  whole energy regime (left) and the region in the gap (right);
  $\Delta_{\rm sc}=0.02$ and  $\pi\Gamma= 0.2$.}    
\label{offdiagspecdel0.02}
\end{figure}

The spectral function of the offdiagonal Green's function at the impurity
site (\ref{offdiagspecfct}) for this value of the gap,
$\Delta_{\rm sc}=0.02$, is shown in figure \ref{offdiagspecdel0.02} for a
number of different values of $U$.     

\noindent 
For larger frequencies outside of the gap (left) we can see a peak near
$\omega=\Delta_{\rm sc}$, whose height is reduced on increasing $U$. 
At larger frequencies we find that the tails develop a broad peak for
larger values of $U$. This has not been observed in the case with the smaller
gap shown in figure \ref{offdiagspecdel0.005}. Also a sign change of the low 
energy peak is found as before. The behaviour near and in the gap
(right) can be understood as before, where in this case
we have shown two values of $U$ with a singlet ground state and two
with a doublet ground state.

\subsubsection{Analysis of bound states with renormalised parameters}
\begin{figure}[!htbp]
\centering
\includegraphics[width=0.45\textwidth]{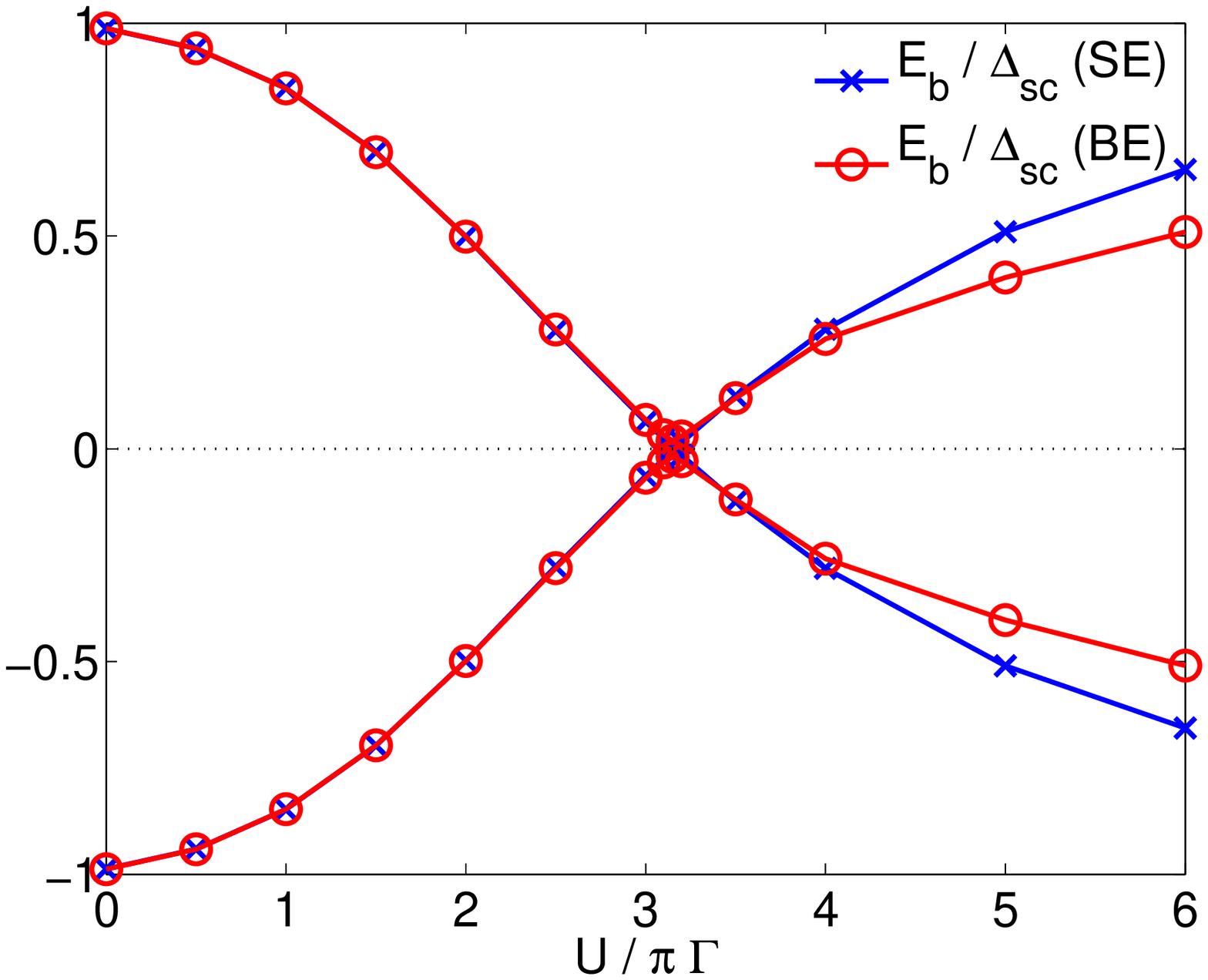}
\hspace{0.5cm}
\includegraphics[width=0.45\textwidth]{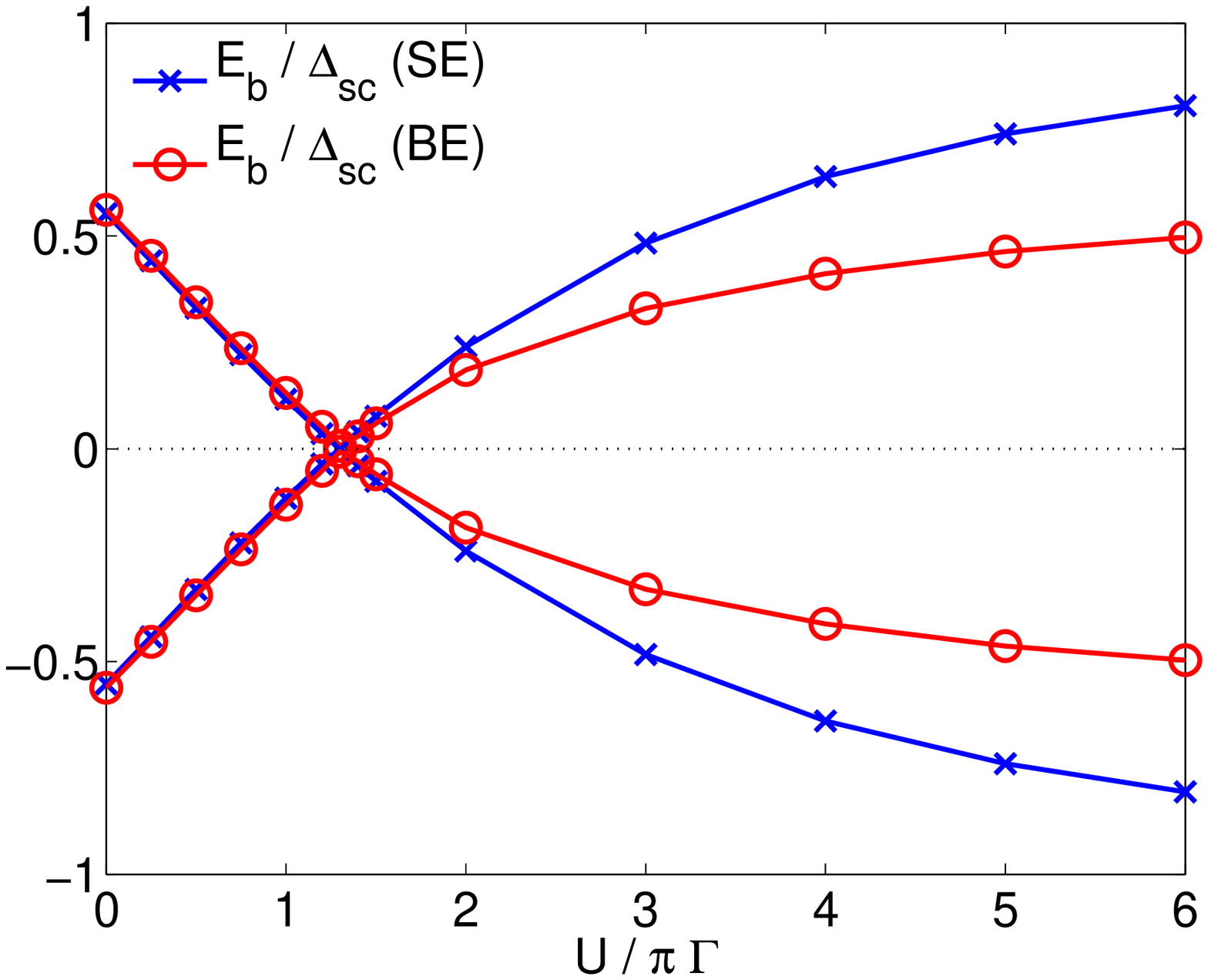}
\vspace*{-0.5cm}
\caption{Bound state energies $E_b$ as calculated from the spectral
  excitations (SE) and from the bound state equation (BE) (\ref{intactBE}) with
  renormalised parameters for $\Delta_{\rm sc}=0.005$ (left) and for
  $\Delta_{\rm sc}=0.06$ (right) for various $U/\pi\Gamma$;
  $\pi\Gamma= 0.2$ is fixed.}     
\label{compboundstates}
\end{figure}

In section \ref{sec:aimsc} we have discussed how the bound state
energy, which so far was deduced from the spectral excitations (SE), can also be
calculated from the bound state equation (BE) (\ref{intactBE}). The latter was
derived by expanding the self-energy to first order. It
involves the renormalised parameters $\tilde\epsilon_d$, $\tilde \Gamma$ and
the constant value of the offdiagonal self-energy  $\Sigma^{\rm off}(0)$. In
figure \ref{compboundstates} we compare the bound state energies calculated by
these two methods for two values of the gap $\Delta_{\rm sc}=0.005$ (left) and
$\Delta_{\rm sc}=0.06$ (right).  

\noindent
We can see that for values of $U<U_c$ the agreement is excellent in both
cases. However, when $U\ge U_c$ we find less accurate values with the method based on
bound state equation (BE) with renormalised parameters. Since the method to
calculate the bound state energy from the NRG spectral excitations (SE) is very
accurate we expect inaccuracies to be found in the BE method. Indeed,
the closer inspection of the numerical results for the diagonal and
off-diagonal self-energies reveals that the linear and constant
approximation made in section \ref{sec:andbs} to derive the bound 
state equation with renormalised parameters (\ref{intactBE}) becomes less
applicable for $U\ge U_c$. The self-energy displays additional features there.

\begin{figure}[!htbp]
\centering
\includegraphics[width=0.45\textwidth]{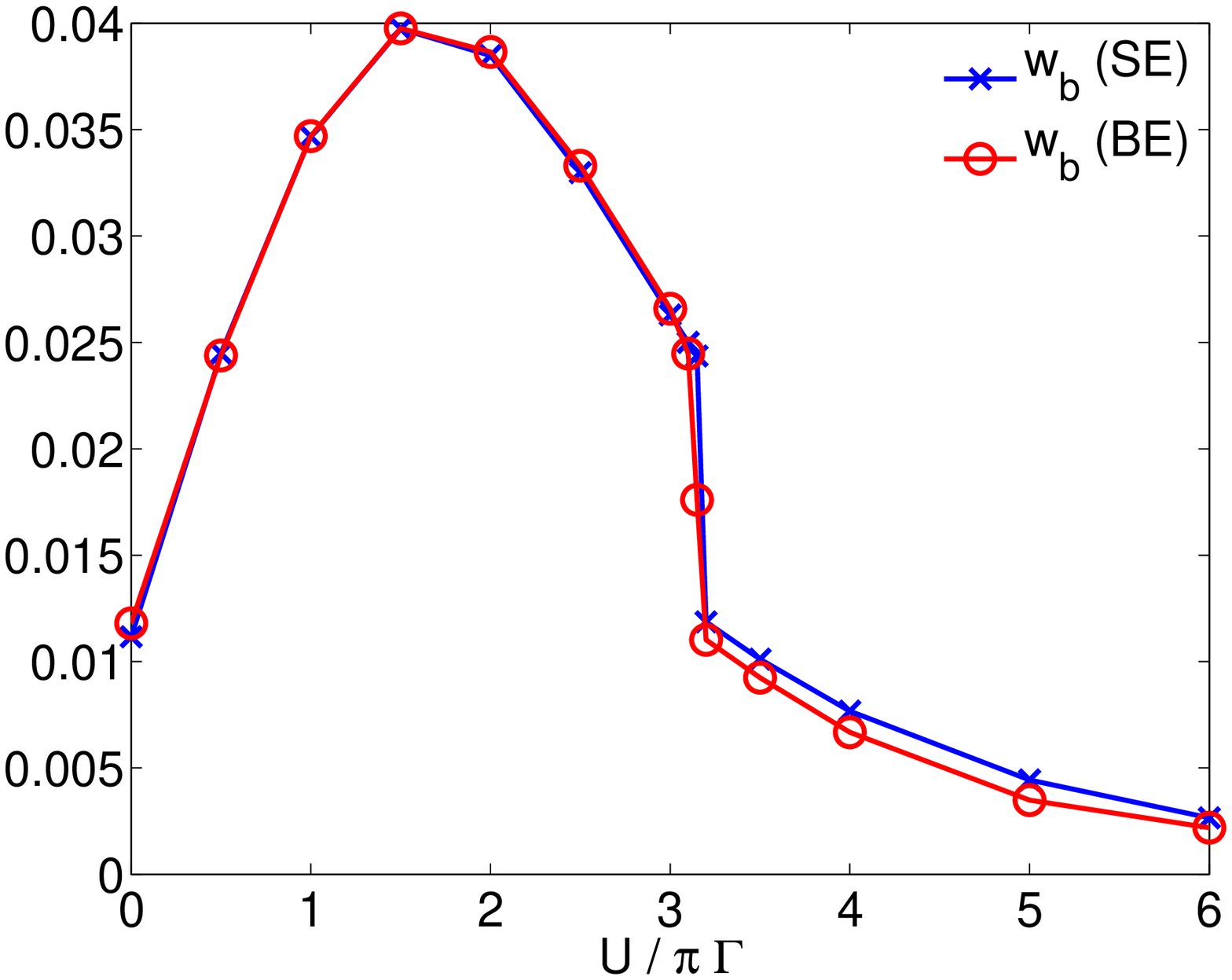}
\hspace{0.5cm}
\includegraphics[width=0.45\textwidth]{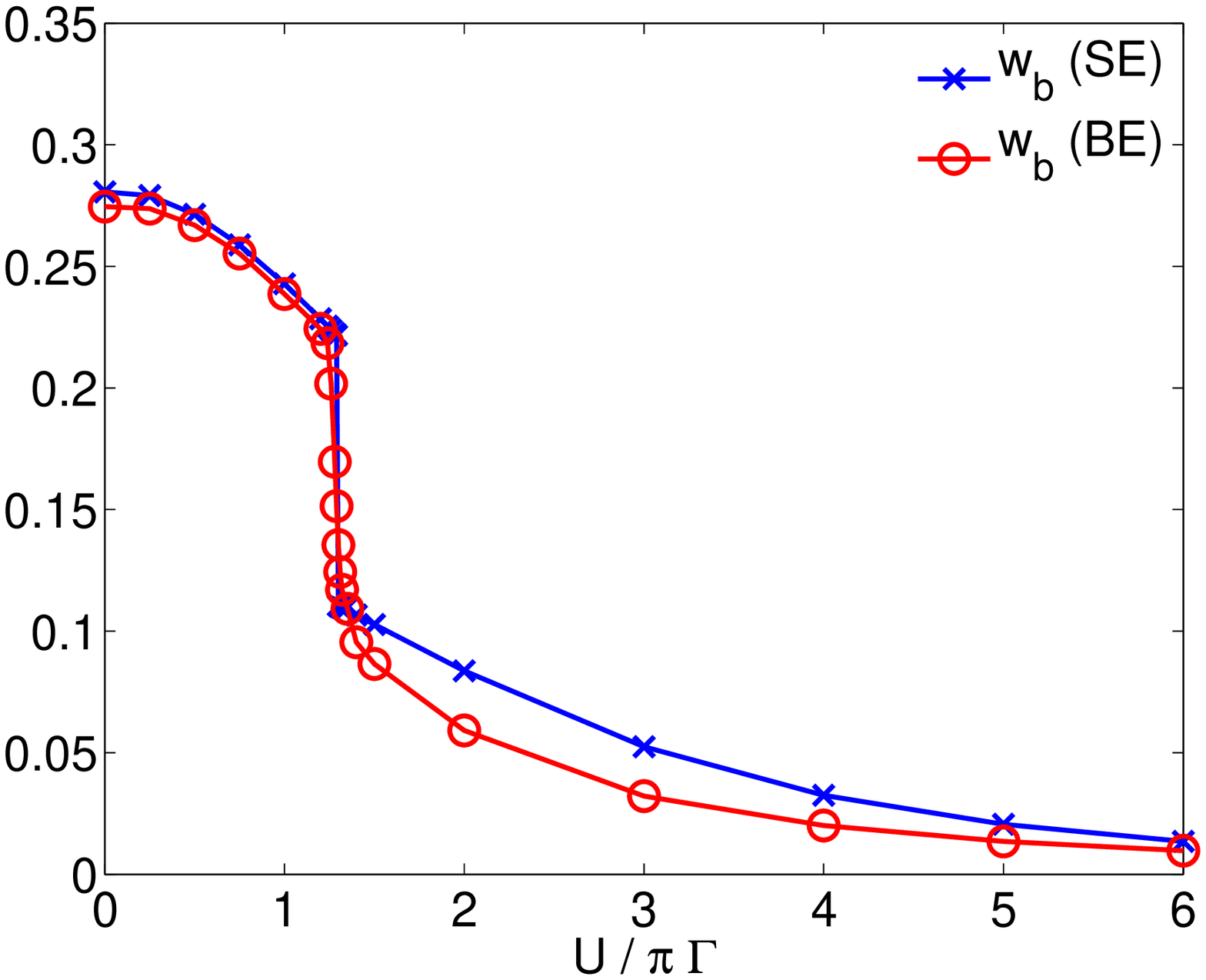}
\vspace*{-0.5cm}
\caption{Weights $w_b$ for the Andreev bound states as calculated from the spectral
  excitations (SE) and from the equation (\ref{weightsbs}) with  renormalised
  parameters for   $\Delta_{\rm sc}=0.005$ (left) and for $\Delta_{\rm
  sc}=0.06$ (right) for various $U/\pi\Gamma$;  $\pi\Gamma= 0.2$ is fixed.}      
\label{compweights}
\end{figure}

In section \ref{sec:aimsc} we have also derived an expression
(\ref{weightsbs}) for the 
weights $w_b$ of the bound states in the gap. It can be expressed in terms of
the renormalised parameters $\tilde\epsilon_d$, $\tilde \Gamma$, the
offdiagonal self-energy  $\Sigma^{\rm off}(0)$ and the bound states energy
$E_b$. In figure \ref{compweights} we compare the weights calculated from
the spectral excitations (SE) with the ones from the bound state equation (BE)
analysis with renormalised parameters.
We show the results for the same parameters $\Delta_{\rm sc}=0.005$ (left) and
$\Delta_{\rm sc}=0.06$ (right).
 
\noindent
We can see for both cases that the overall
behaviour of the weights as a function of $U$ is described reasonably well by
equation (\ref{weightsbs}). It is, however, clearly visible that the agreement is
between the SE and BE values is much better in the singlet regime for
$U<U_c$. This is similar as 
observed for the values of the bound states energies $E_b$ in figure
\ref{compboundstates}, and the reason for this is the same. The discontinuity
for the weight is not reproduced by the approximation based on equation
(\ref{weightsbs}). As can be seen from that equation this would require a
sudden change in the self-energy as function of $U$, which was not found with
sufficient  accuracy in the present calculation. This can partly be attributed
to the broadening procedure involved and to the inaccuracies when calculating
the numerical derivative.

\subsubsection{Anomalous expectation value and phase diagram}
The anomalous expectation value $\langle d_{\uparrow}d_{\downarrow}\rangle$ is
an indicator for the strength of the proximity effect of the superconducting
medium at the impurity site and quantifies the induced on-site superconducting
correlations. In the following figure \ref{ddexp} we show the dependence of
$\langle d_{\uparrow}d_{\downarrow}\rangle$ on the interaction $U/\pi\Gamma$
for the same values of $\Delta_{\rm sc}$ as in figure \ref{boundstates}.  
The values are scaled by the gap $\Delta_{\rm sc}$. 

\begin{figure}[!htbp]
\centering
\includegraphics[width=0.45\textwidth]{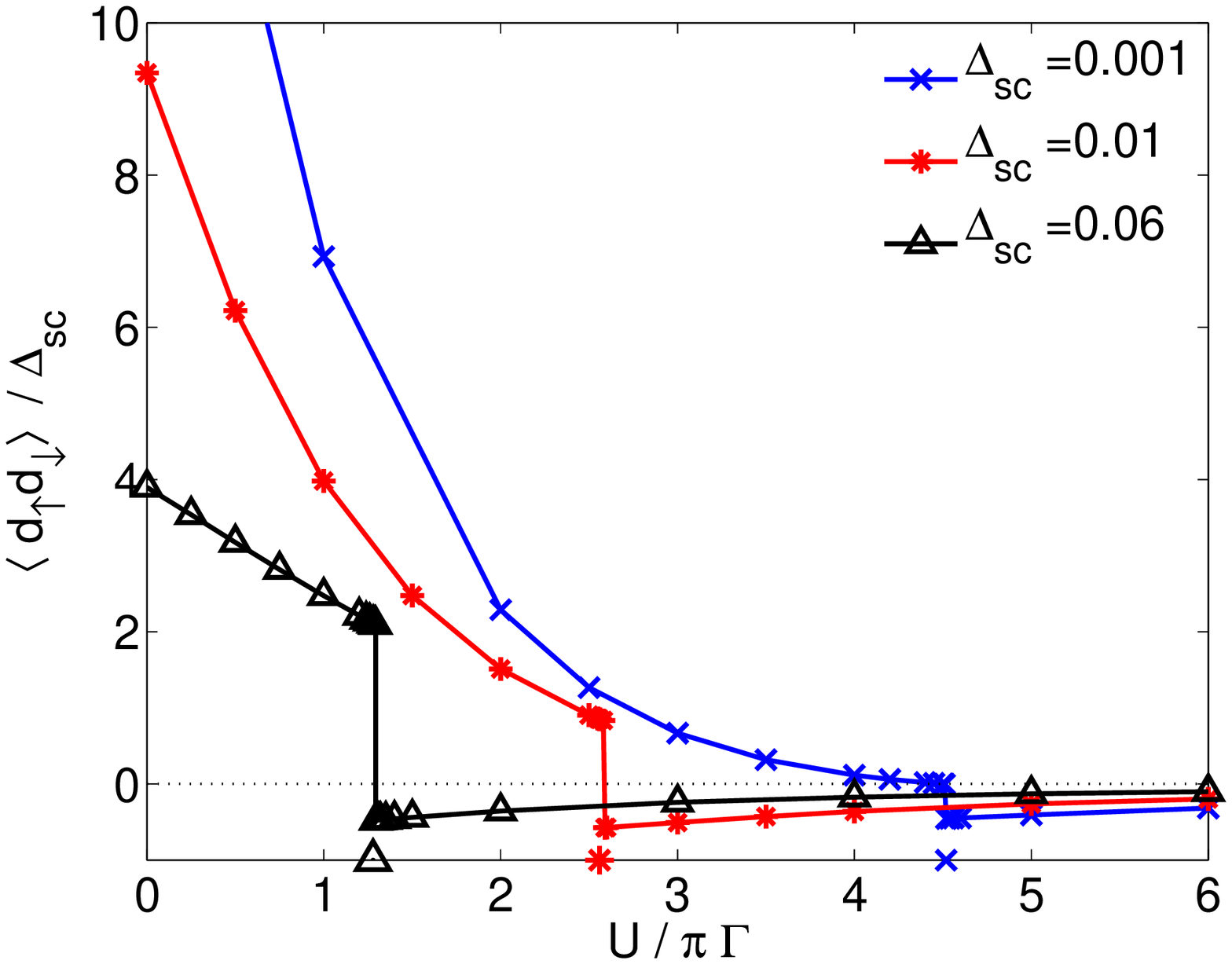}
\hspace{0.5cm}
\includegraphics[width=0.45\textwidth]{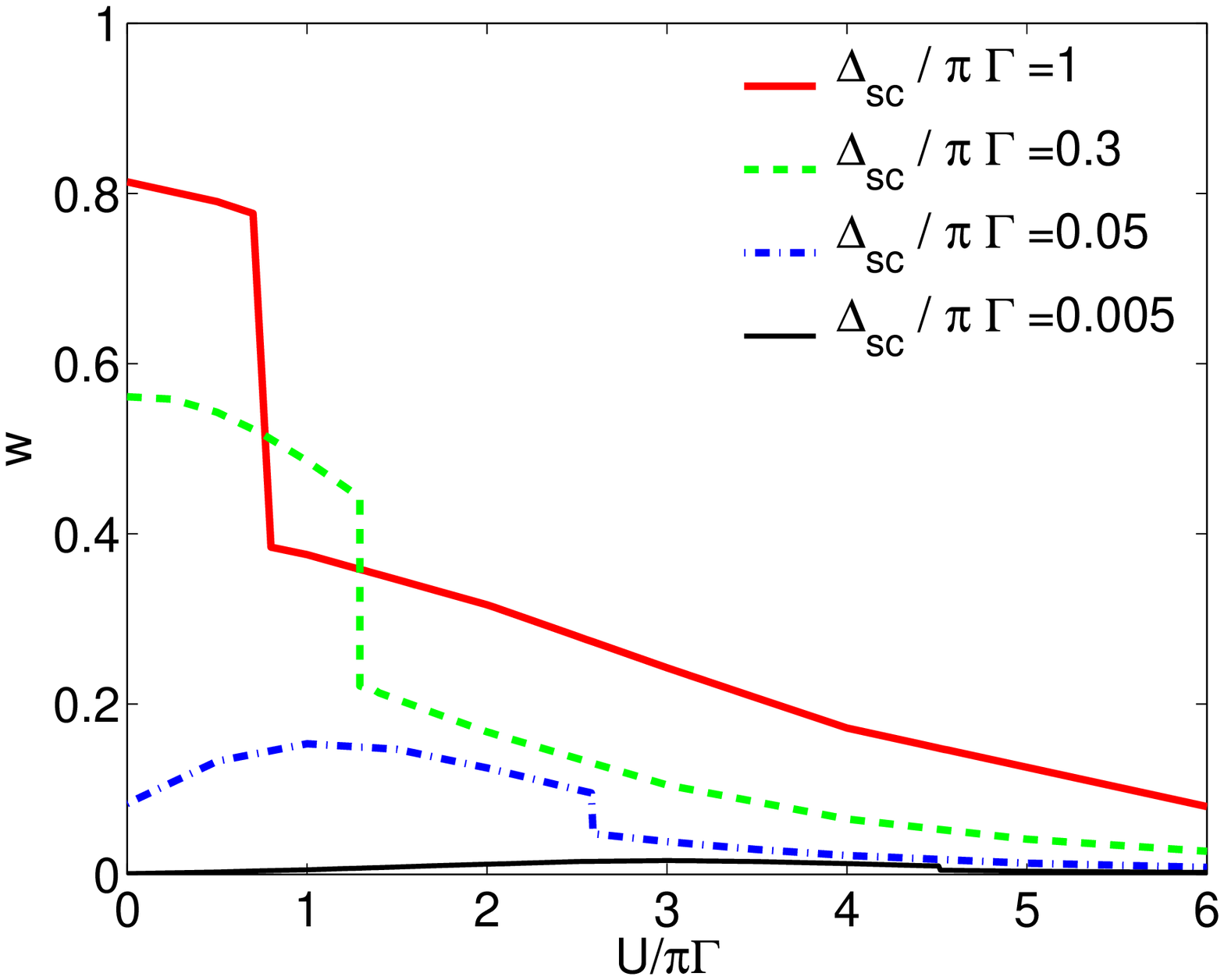}
\vspace*{-0.5cm}
\caption{Left: Anomalous expectation values $\langle
  d_{\uparrow}d_{\downarrow}\rangle$ as a function of $U/\pi \Gamma$ 
  for various $\Delta_{\rm sc}$. The values are scaled by the gap $\Delta_{\rm
  sc}$;  $\pi\Gamma= 0.2$.  Right: The total weight of the bound
states $w$ in the gap as calculated from the spectral excitations as a
function of $U/\pi\Gamma$ for various $\Delta_{\rm sc}/\pi\Gamma$.}      
\label{ddexp}
\end{figure}
\noindent
We see that as a general trend $\langle d_{\uparrow}d_{\downarrow}\rangle$
decreases for increasing on-site interaction. This is 
expected since the superconducting correlations are suppressed by the
repulsive interaction. We have marked the ground state transition with a
symbol on the $x$-axis, and we see that $\langle
d_{\uparrow}d_{\downarrow}\rangle$ changes discontinuously in magnitude and 
sign there.  This is characteristic for this zero temperature quantum phase
transition. The sign change is due to a phase change of $\pi$ of
the local order parameter which occurs at the transition as discussed
in reference \cite{BVZ06}.
In the situation of infinite gap in the medium,
which was discussed in section \ref{sec:limlargegap},  $\langle
d_{\uparrow}d_{\downarrow}\rangle$ drops only to zero at the
transition point and is zero in the doublet ground state. At finite
temperature the behaviour becomes continuous.

An overview of the transfer of spectral weight from the continuum to the bound
states is shown in figure \ref{ddexp} (right). There
we plot the total weight $w=w_b^+ +w_b^-$ as a function of $U/\pi\Gamma$
for four selected values of $\Delta_{\rm sc}/\pi\Gamma$ ranging from
0.005 to 1. The curves are similar as before in figure
\ref{boundstates} and show the discontinuity at the ground state
transition. Here the values are not scaled by $\Delta_{\rm sc}$. We
can see that the smaller $U$ and the larger $\Delta_{\rm sc}$ are
the more spectral weight is found in the bound states. In the extreme
case of $\Delta_{\rm sc}\to 0$  we have $w=0$, 
and for large gap, $\Delta_{\rm sc}\to \infty$, and small $U$
equation (\ref{weightsbs}) gives $w\to 1$. The tendency to both of
these limiting cases can be inferred from figure \ref{ddexp} (right)
and we can see that, for instance, for $\Delta_{\rm sc}=\pi\Gamma$
already about 80\% of the spectral weight is in the bound states.

Summarising the behaviour for different parameters, we present a phase
diagram for  singlet and doublet states for the symmetric model in
the following figure \ref{hfphasedia}.

\begin{figure}[!htbp]
\centering
\includegraphics[width=0.65\textwidth]{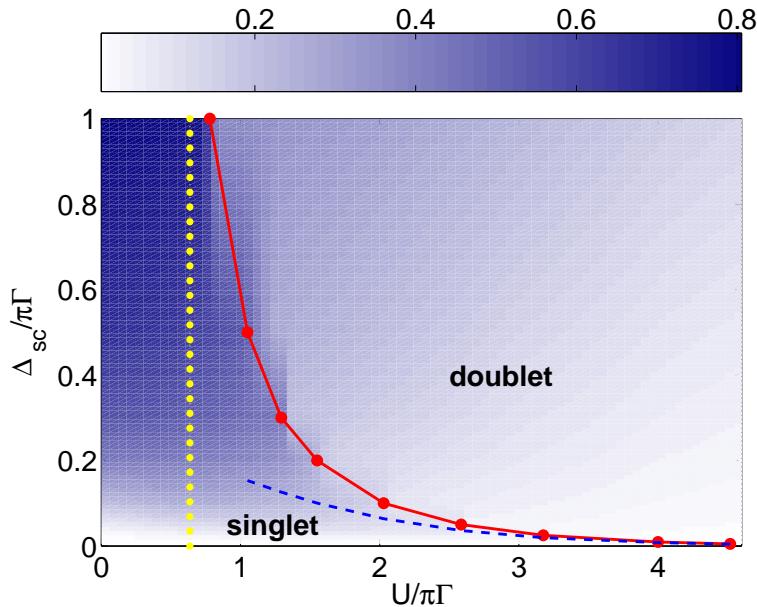}
\vspace*{-0.5cm}
\caption{Phase diagram for singlet and doublet ground-state
  as a function of $\Delta_{\rm sc}/\pi \Gamma$ and $U/\pi \Gamma$,
  where the full line with large dots describes the phase boundary. The dotted
  line corresponds to $U/\Gamma=2$, which shows the singlet doublet transition
  for $\Delta_{\rm sc}\to\infty$. The dashed line gives the transition as
  $T_{\rm  K}/\Delta_{\rm sc}\simeq 0.3$ with $T_{\rm  K}$  given in equation
  (\ref{yoshiokatk}). As a background colour we have included the
  amount of spectral weight transfered to the bound states; the
  discontinuous behaviour at the the singlet
  doublet ground state transitions is slightly blurred in the
  interpolated representation.}      
\label{hfphasedia}
\end{figure}
\noindent
For small $U$ the ground
state is always a singlet. It can become a doublet when $U/\pi \Gamma$ is
increased. The critical $U_c$ for the transition decreases with increasing
value of the gap $\Delta_{\rm sc}$ as can be seen in the diagram. In the limit
$\Delta_{\rm sc}\to \infty$, the critical interaction is given by
$U_c/\pi\Gamma=2/\pi$, which is shown with a dotted vertical
line in the figure. As mentioned in the Introduction there have been
estimates of the phase boundary for the singlet and doublet ground
state in the strong coupling regime \cite{SSSS92,YO00} as $T_{\rm
  K}/\Delta_{\rm   sc}\simeq 0.3$. In this case the Kondo temperature
is given as in equation (3.9) in reference \cite{YO00},   
\begin{equation}
  \label{yoshiokatk}
T_{\rm K}=0.182 U\sqrt{\frac{8\Gamma}{\pi U}}\e^{-\pi U/8\Gamma}.
\end{equation}
We have added a dashed line representing this result which agrees with the
ones presented here in the strong coupling regime, but starts to deviate for
smaller values of $U$. As a background colour we have included in
figure \ref{hfphasedia} how much spectral weight $w$ is transfered to the
bound states (The value of $w$ is given by the colour bar on the top part of
the figure.). As noted before in figure \ref{ddexp} (right) we can see
generally that the weight is maximal in the region of large gap and
small on-site repulsion $U$.

\begin{figure}[!htbp]
\centering
\includegraphics[width=0.45\textwidth]{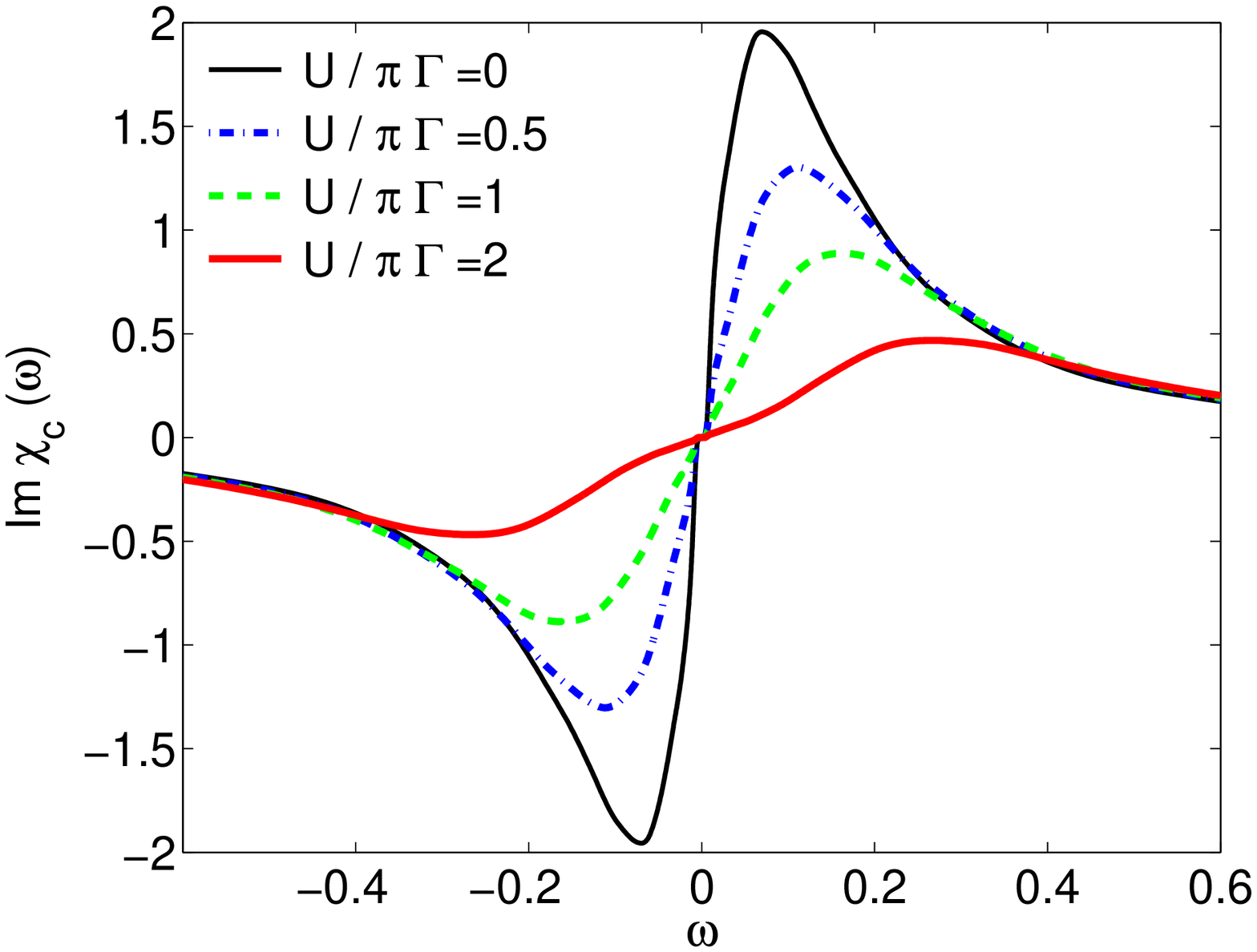}
\hspace{0.5cm}
\includegraphics[width=0.45\textwidth]{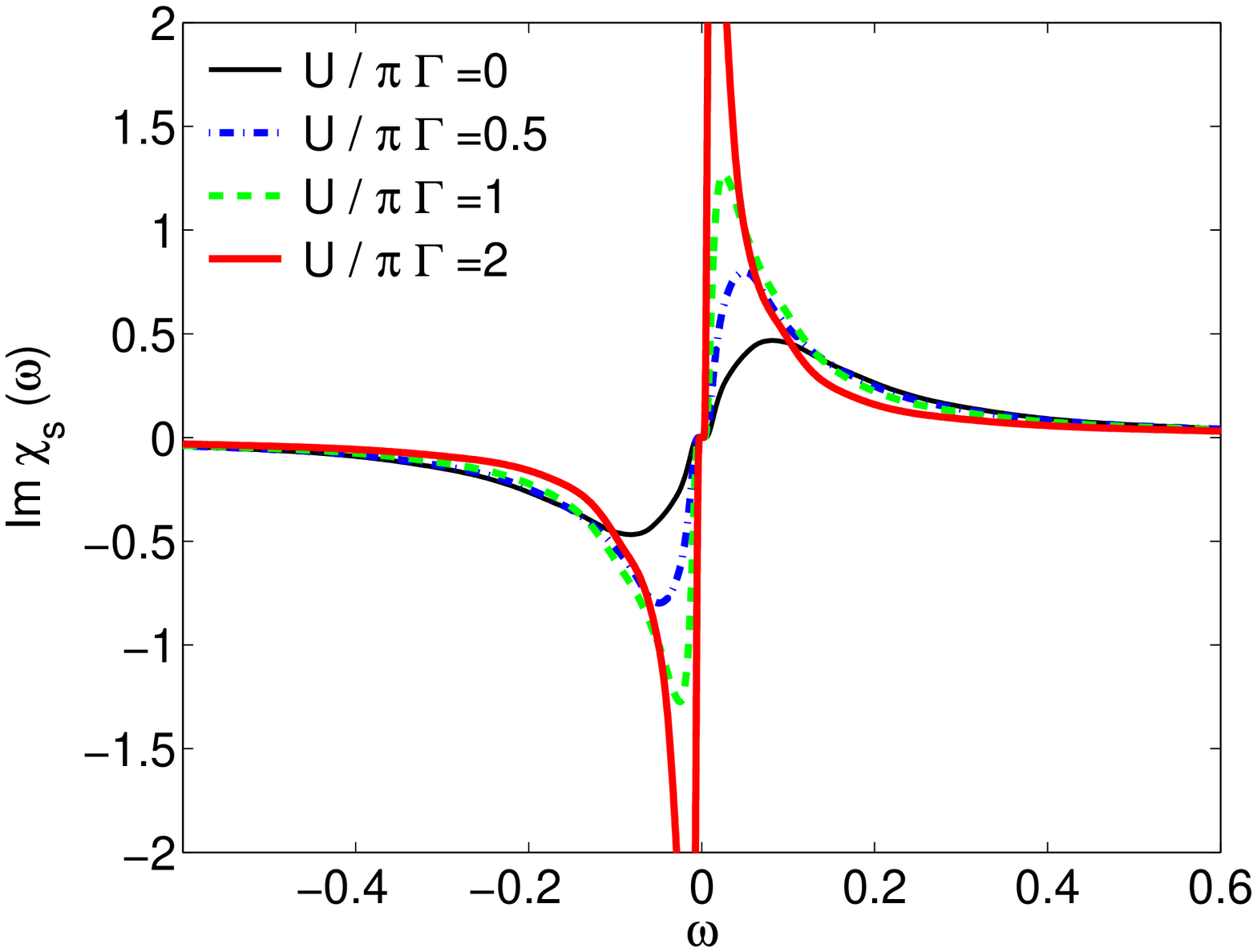}
\vspace*{-0.5cm}
\caption{The imaginary  part of the dynamic charge (left) and spin (right)
  susceptibility  various values of $U$; $\Delta_{\rm sc}=0.005$ and
  $\pi\Gamma= 0.2$. The scale on both axes is the same such that the
  results can be compared well.}     
\label{suscdel0.005}
\end{figure}

At $\Delta_{\rm sc}\to 0$ the ground-state is a singlet for
any value of $U$ as the Kondo effect always leads to a screened
impurity spin in a singlet formation. For finite gap the nature of the
singlet ground state can differ depending on the magnitude of $U$. 
It can be a singlet corresponding to an s-wave pair like in the wave  
function given in equation (\ref{wffct}), which is a superposition of zero and
double occupation. This is the natural singlet ground state for a BCS
superconductor. In the strong coupling regime 
we can, however, also have a screened local spin, i.e. a Kondo
singlet. The wave function has a 
different form then and consists rather of a singly occupied impurity state
coupled to the spins of the medium as many-body state. In our NRG calculations it
is not easy to distinguish clearly this different nature of the
singlet ground states and draw a definite line to separate them. We
can, however, get an indication for what is favoured from the two
particle response functions in the spin and in the charge channel.  In
figure \ref{suscdel0.005} we show the imaginary  part of the dynamic
charge and spin susceptibility, $\chi_c(\omega)$ and $\chi_s(\omega)$,
for $\Delta_{\rm sc}=0.005$ and a series of values for the interaction
$U$. 

\noindent 
We can see that the peaks in the charge susceptibility exceed the ones in the
spin susceptibility for zero and weak interaction indicating the dominance
of the symmetry breaking in the charge channel, and a ground state of
superconducting singlet nature. However, for $U/\pi\Gamma>1$ the
spin susceptibility develops a large and narrow peak at low frequency. This signals
the importance of the spin fluctuations and low energy spin excitations and
indicates a ground states of a screened spin. In contrast the decreasing peaks in
the charge susceptibility for large $U$ is consistent with  the effect
of suppression of the on-site superconducting correlations.

\subsection{Away from particle-hole symmetry}
\label{sec:scawphsym}
So far we have considered the special situation of particle-hole symmetry,
$\epsilon_d=-U/2$. In this section we will briefly discuss a few aspects
that change in the situation away from particle-hole symmetry.
Let us consider the case where for a given gap $\Delta_{\rm sc}$,
on-site interaction $U$, and hybridisation $\Gamma$, the ground-state of the system is a
doublet at half filling, $\xi_d=0$. When $\xi_d$ is increased, we find that a
transition to a singlet state can occur at a certain value $\xi_d^c$. 
This is illustrated
in the following figure \ref{boundstatesasy}, where we have plotted the bound
state energy $E_b$ for fixed $\Delta_{\rm sc}=0.01$, two values of
$U/\pi\Gamma=3,5$ and a series of values of the on-site energy scaled by $U$, $\xi_d/U$. 
As before we have $\pi\Gamma= 0.2$.

\begin{figure}[!htbp]
\centering
\includegraphics[width=0.47\textwidth]{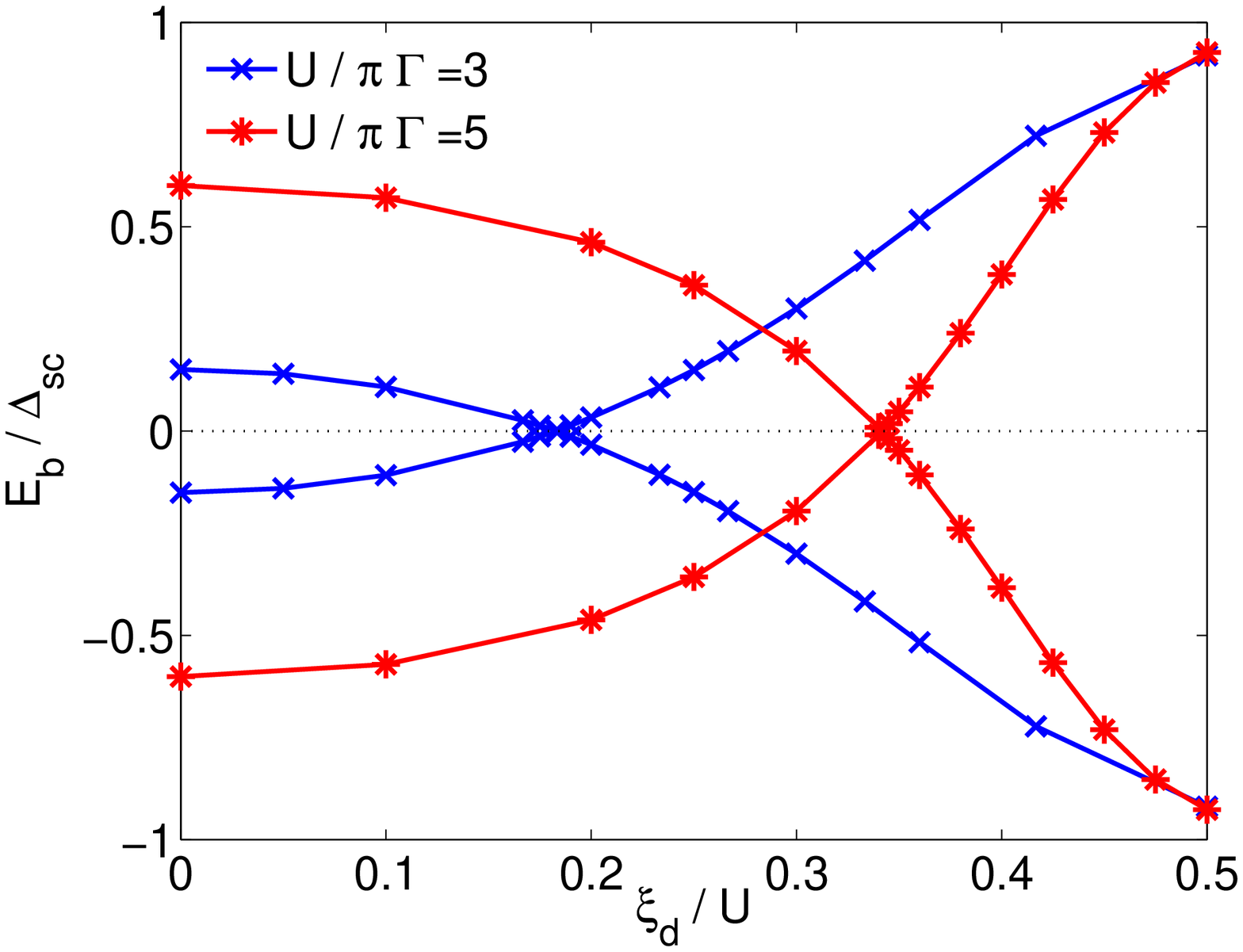}
\hspace{0.5cm}
\includegraphics[width=0.45\textwidth]{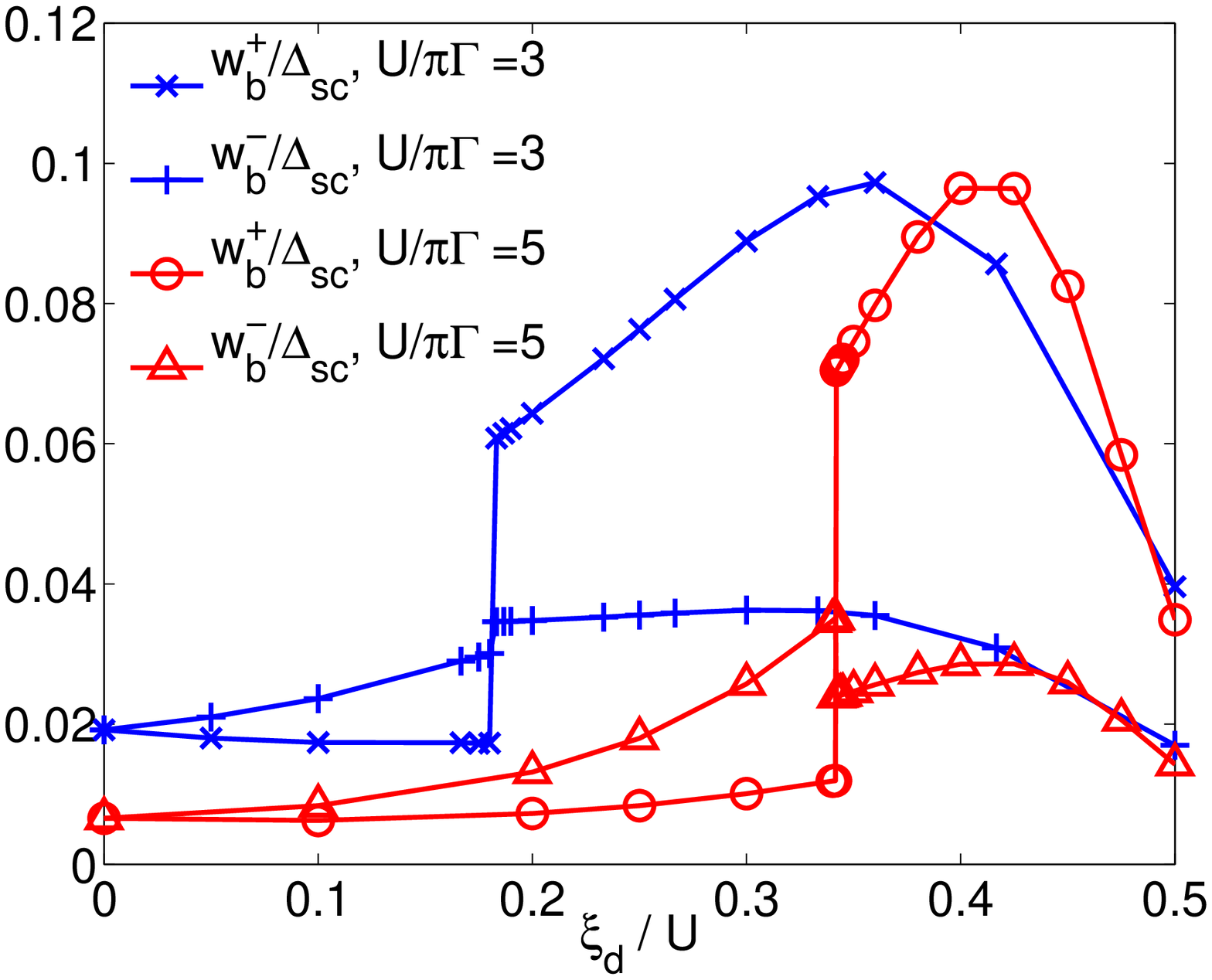}
\vspace*{-0.5cm}
\caption{The dependence of the bound state energies $E_b$ (left) and weights $w_b$ (right)
   on $\xi_d/U$ for $\Delta_{\rm sc}=0.01$ and  $U/\pi\Gamma=3,5$;
   $\pi\Gamma= 0.2$ is fixed.}      
\label{boundstatesasy}
\end{figure}
\noindent
The critical interaction for the ground state transition for this case at half
filling is $U_c/\pi\Gamma\simeq 2.6$, such that both cases possess a doublet
ground state for $\xi_d=0$. We can see that with increasing asymmetry $\xi_d$
the bound state energy $\mod{E_b}$ first decreases towards zero and then increases again
in the singlet regime for $\xi_d>\xi_d^c$. As in the symmetric case the
singlet-doublet transition is accompanied by $\mod{E_b}=0$. 
The weights $w_b^{\pm}$ for these bound states are shown on the right hand side of figure
\ref{boundstatesasy}. Away from particle-hole symmetry the weight $w_b^+$ for
the positive energy $E_b^+$ and $w_b^-$ the one for the negative bound state
$E_b^-$ are not equal, as was already pointed out below equation (\ref{weightsbs}). 
We can see that the weights $w_b^{\pm}$ start to assume different values when $\xi_d$ is
increased from 0. At the ground state transition the values change
discontinuously similar as observed in the half filled case.
If we follow  both the positive weight $w_b^+$ and the negative
$w_b^-$ separately the weights cross at the transition point. If, however, we
think of the  bound states as crossing at zero, i.e. $w_b^+\leftrightarrow
w_b^-$ at the transition, a more direct connection can be deduced
from the results shown. In the singlet phase there is a maximum for both the
positive and the negative bound state weight, more pronounced for $w_b^+$.

Also in the asymmetric case it is possible to calculate the bound state
position $E_b$ from equation (\ref{iaandbs}) and the weights from equation
(\ref{weightsbs}) employing the renormalised parameters. We do not show the  plots here, but note that the results resemble figures
\ref{compboundstates} and \ref{compweights} in the respect that they give good
agreement in the singlet regime, but deviations for parameters where the ground
state is a doublet.

In the following figure \ref{phasediag} (left) we show the dependence of the anomalous
expectation value $\langle d_{\uparrow}d_{\downarrow}\rangle$ on the
asymmetry scaled by the interaction $\xi_d/U$ for the same value of $\Delta_{\rm sc}$ as in
figure \ref{boundstatesasy}. 
\begin{figure}[!htbp]
\centering
\includegraphics[width=0.45\textwidth]{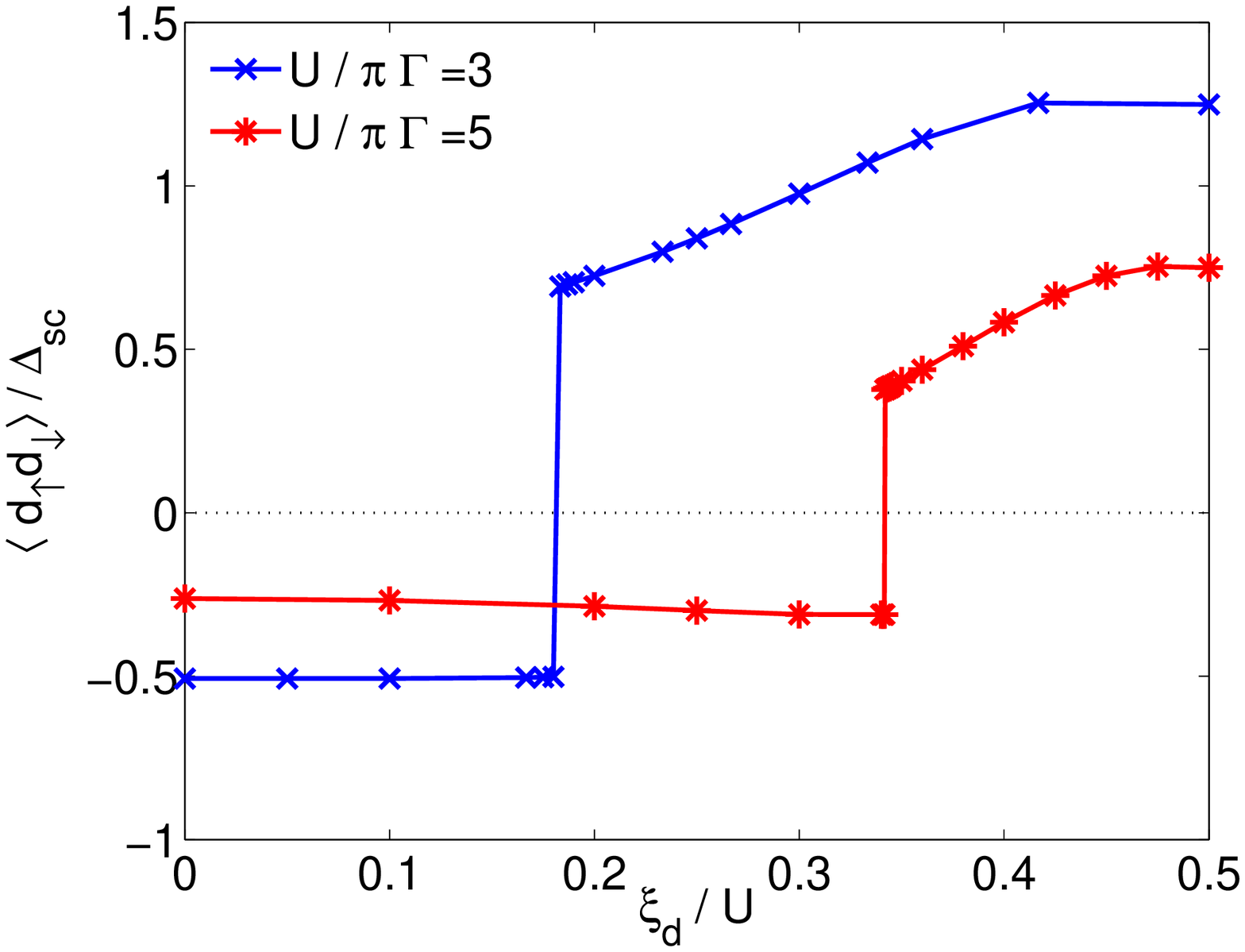}
\hspace{0.5cm}
\includegraphics[width=0.45\textwidth]{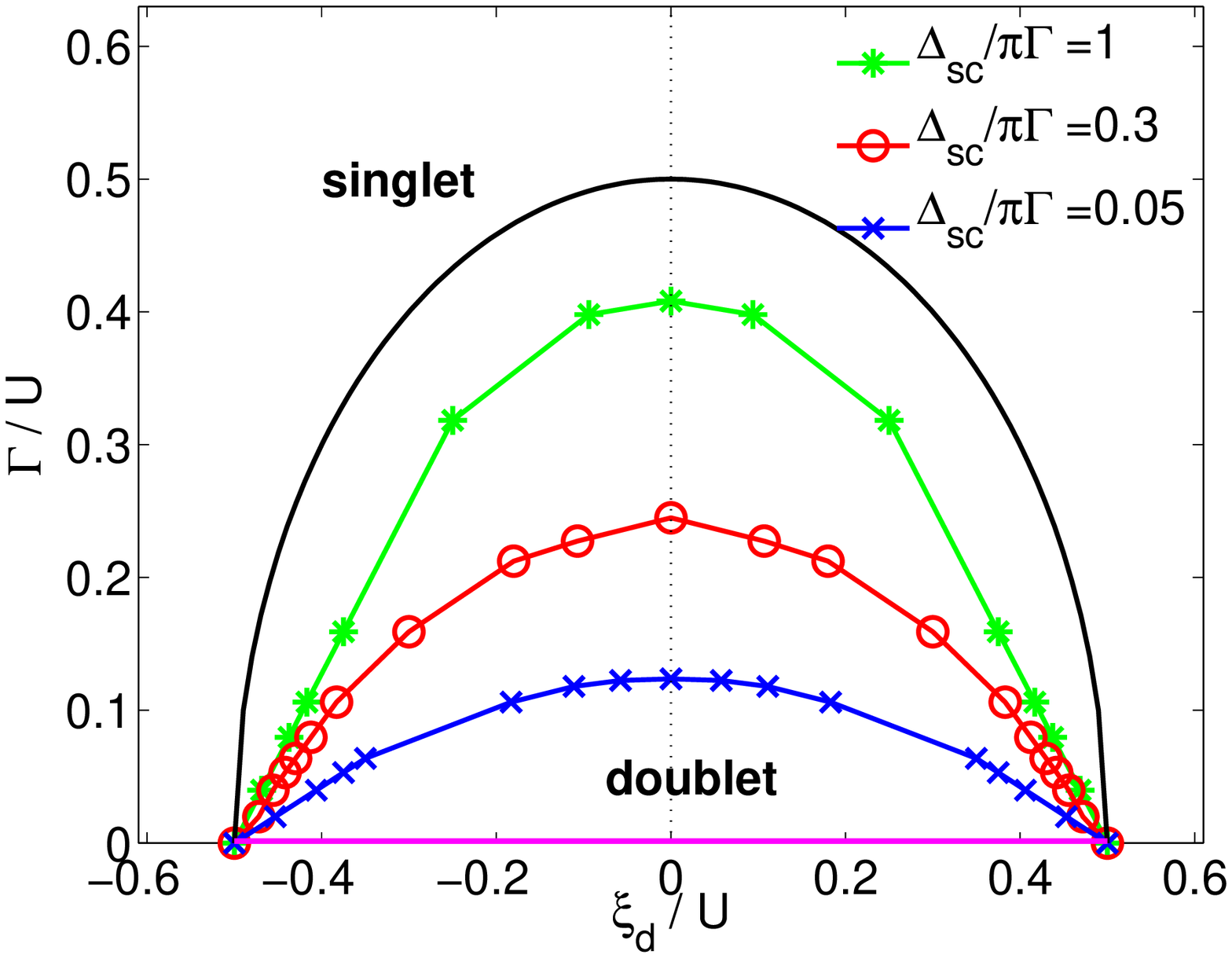}
\caption{Left: Anomalous expectation values  $\langle
  d_{\uparrow}d_{\downarrow}\rangle$ for various $U/\pi \Gamma$,
  $\Delta_{\rm sc}=0.01$ and  $\pi\Gamma= 0.2$. Right: Phase diagram
  showing the regions for singlet and doublet ground 
  state as dependent on $\Gamma/U$ and $\xi_d/U$ for different values of the
  gap $\Delta_{\rm sc}$. The full semicircular line corresponds to the
  phase boundary for $\Delta_{\rm sc}=\infty$ as discussed in equation
  (\ref{delinfphasbound}).}      
\label{phasediag}
\end{figure}
\noindent 
The values for $\langle d_{\uparrow}d_{\downarrow}\rangle$ are scaled by the
gap $\Delta_{\rm sc}$. For the values of $U$ shown, at half filling the system
has a doublet ground state and $\langle d_{\uparrow}d_{\downarrow}\rangle$ 
is negative. First it does not vary much when $\xi_d$ is increased, but at the 
transition to the singlet ground state we find, as in the half filled case, a
jump to a positive value and $\langle d_{\uparrow}d_{\downarrow}\rangle$
increases to a saturation value on further  increasing $\xi_d$. This value is
smaller for larger $U$, similar to what has been found in the symmetric case.

On the right hand side of figure \ref{phasediag} we present a global phase
diagram of the parameter regimes for singlet and doublet ground states for the
non-symmetric case. This 
representation in the $\Gamma/U$-$\xi_d/U$-plane is motivated by the result
for the phase boundary for the case $\Delta_{\rm sc}\to \infty$
derived in section \ref{sec:limlargegap}, equation
(\ref{delinfphasbound}). The  semicircle corresponding to this case is shown in the 
figure together with the phase boundaries for some finite values of the gap
$\Delta_{\rm sc}$. These are seen to have a similar form, but the
boundary decreases to smaller values of $\Gamma/U$ with $\Delta_{\rm
  sc}/\pi\Gamma$. Note that the parameters on the line on 
the $x$-axis, to which the phase boundary contracts  in the limit
$\Gamma\to 0$ or $U\to\infty$, possess a doublet ground state for
$\mod{\xi_d}/U<1/2$.


\section{Conclusions}

We have discussed and quantitatively analysed the different forms of
behaviour that can occur for an interacting impurity  site in a medium
with offdiagonal symmetry breaking in the charge channel. This study is 
motivated by the experimental situations of impurities in superconductors and
nanoscale quantum dot systems with superconducting leads. 
In the local spectral functions we found that the low energy spectrum
is  dominated by the superconducting gap, and we saw that the lowest excitations in
these cases are Andreev bound states within the gap region. For higher
energies the spectrum resembles the form usually found in a metallic
bath with broadened atomic limit peaks for large $U/\pi\Gamma$. The
formation of the Kondo resonance, whose width is proportional to
$T_{\rm K}$, is in direct competition with the superconducting 
spectral gap of magnitude $\Delta_{\rm sc}$. Therefore, depending on
the ratio of these parameters a screened Kondo singlet or an unscreened
local moment is observed.

The lowest spectral excitations, the Andreev bound states within the
gap region, change position and weight according to the other
parameters. These have been analysed in detail in both the symmetric
and the asymmetric model. We have given a simple interpretation of
their position and weight in terms of renormalised parameters. It
turned out that the assumptions for the definition of these were satisfied
better in the singlet ground state regime. The reason for this should be 
subject of further investigation. 
In the quantum dot systems currents have been observed involving
multiple Andreev processes \cite{BNS02,BBNBBS03}. It is expected that
a quantitative understanding of these currents require accurate
information about the weight and position of the Andreev bound
states, which have been provided here. To study the experimental
situation in detail and to describe the differential conductance
dependent on the local bound state behaviour can be subject of a 
separate publication, where also the details of the experimental setup
are taken into account more carefully. 

The behaviour of the ground state of the system, which can be a
spin singlet or a doublet, is summarised in the two phase 
diagrams in figures \ref{hfphasedia} and \ref{phasediag}. For the
overlapping parameter ranges our results for the ground state and the
locally excited states are in  agreement with earlier NRG studies 
\cite{SSSS92,SSSS93,YO00}. Differences can be seen in the spectral 
representation of the bound states in the gap. Here we report delta
function peaks, whereas an earlier study \cite{CLKB04} presented
broadened peaks. 
The method of calculating spectral functions and the self-energy used
and explained in the appendix of this paper will be relevant for
extensions of the calculation to the lattice model within the
dynamical mean field theory framework. There an effective Anderson
impurity model could be used to study the phases with superconducting
symmetry breaking for instance in the attractive Hubbard model.

 \bigskip\par
  
\ack

We would like to acknowledge helpful discussions with R. Bulla and
Hyun-Jung Lee. JB is grateful for the hospitality at Osaka City
University, where this work was initiated, and to SFB 484 at the
University of Augsburg, where this work was finalised. 
JB thanks the Gottlieb Daimler- and Karl Benz-Foundation, EPSRC, the
DAAD and JSPS for financial support, and AO acknowledges the support
by the Grant-in-Aid for Scientific Research for JSPS. We also wish to
thank W. Koller and D. Meyer for their earlier contributions to the
NRG program.  

\par
\begin{appendix}
\section{Relevant Green's functions}
For the Green's functions it is convenient to work in Nambu space,
$\vct C^{\dagger}_{d}=(\elcre{d}{\uparrow},\elann{d}{\downarrow})$,
with
$2\times2$ matrices.
The relevant retarded Green's functions are then 
\begin{equation}
\fl
  \underline {G}_d(\omega)=
\gfbraket{\vct C_{d};\vct C^{\dagger}_{d}}_{\omega}=
\left(\begin{array}{c c}
\gfbraket{\elann{d}{\uparrow};\elcre{d}{\uparrow}}_{\omega} & 
\gfbraket{\elann{d}{\uparrow};\elann{d}{\downarrow}}_{\omega} \\
\gfbraket{\elcre{d}{\downarrow};\elcre{d}{\uparrow}}_{\omega} &
\gfbraket{\elcre{d}{\downarrow};\elann{d}{\downarrow}}_{\omega}
\end{array}\right)
=\left(\begin{array}{c c}
G_{11}(\omega) & G_{12}(\omega) \\
G_{21}(\omega) & G_{22}(\omega)
\end{array}\right).
\end{equation}
In the NRG approach we calculate $G_{11}$ and $G_{21}$ directly and infer
$G_{22}(\omega)=-G_{11}(-\omega)^*$,
which follows from $G_{A,B}^{\rm ret}(\omega)=-G_{B,A}^{\rm adv}(-\omega)$ and
$G_{A,B}^{\rm ret/adv}(\omega)=-G_{A^{\dagger},B^{\dagger}}^{\rm
  ret/adv}(-\omega)^*$ for fermionic operators $A$, $B$. Similarly, we can  find
$G_{12}(\omega)=G_{21}(-\omega)^*$. In the derivation one has to be careful and include
a sign change for up down spin interchange in the corresponding operator
combination. 

In the non-interacting case we can deduce the $d$-site Green's function matrix exactly.
To do so rewrite the term $H_{\rm sc}$ by introducing the vector of operators and
the symmetric matrix
\begin{equation}
  \label{cknambu}
\vct C_{\vk}:=
\left(\begin{array}{c}
\! \elann{\vk}{\uparrow}  \\
\! \elcre{-\vk}{\downarrow}
\end{array}\right),
\qquad
A_{\vk}:=
 \left(\begin{array}{cc}
\! \epsilon_{\vk}     & \! -\Delta_{\rm sc}\\
\! -\Delta_{\rm sc} & \! -\epsilon_{\vk}
\end{array}\right).
\end{equation}
Then $H_{\rm sc}$ can be written as
\begin{equation}
H_{\rm sc}=\sum_{\vk}\vct C_{\vk}^{\dagger}A_{\vk}\vct C_{\vk}.
\end{equation}
The matrix Green's function in the superconducting lead is then given by 
$\underline {g}_{\vk}(i\omega_n)=(i\omega_n \unitop_2-A_{\vk})^{-1}$,
\begin{equation}
 \underline {g}_{\vk}(i\omega_n)^{-1}
 =i\omega_n\unitop_{2}-\epsilon_{\vk}\tau_3
+\Delta_{\rm sc}\tau_1,
\label{freescbandgfct}
\end{equation}
where $\tau_i$ are Pauli matrices.
It follows that
\begin{equation}
 \underline {g}_{\vk}(i\omega_n)
 =\frac{i\omega_n\unitop_{2}+\epsilon_{\vk}\tau_3
-\Delta_{\rm sc}\tau_1}{(i\omega_n)^2-(\epsilon_{\vk}^2+\Delta_{\rm sc}^2)}.
\end{equation}
In the wide band limit with a constant density of states the hybridisation
term takes  the form 
\begin{equation}
  V^2 \frac1N\sum_{\vk}\underline {g}_{\vk}(i\omega_n)=
-\Gamma\frac{i\omega_n\unitop_{2}+\Delta_{\rm sc}\tau_1}
{E(i\omega_n)}.
\end{equation}
We are mostly interested in the limit of zero temperature
 here, and the function in the denominator
$E(z)$ after analytic continuation reads
\begin{equation}
\label{funcEom}
E(\omega)=
\left\{
\begin{array}{c c}
-i{\rm sgn}(\omega)\sqrt{\omega^2-\Delta_{\rm sc}^2}
& {\rm for} \;  |\omega|>\Delta_{\rm sc} \\
\sqrt{\Delta_{\rm sc}^2-\omega^2} 
& {\rm for}\; |\omega|<\Delta_{\rm sc} 
\end{array}\right . .
\end{equation}
In the non-interacting case for $T=0$, we have therefore
\begin{equation}
  \underline {G}^0_d(\omega)^{-1}=\omega\unitop_{2}-\epsilon_{d}\tau_3+
\Gamma\frac{\omega\unitop_{2}+\Delta_{\rm sc}\tau_1}
{E(\omega)}.
\end{equation}
The Green's function is obtained by matrix inversion, which yields 
\begin{equation}
  \underline {G}^0_d(\omega)=
\frac{1}{D(\omega)}
\Big[\omega\Big(1+\frac{\Gamma}{E(\omega)}\Big)\unitop_{2}
-\frac{\Gamma\Delta_{\rm sc}}{E(\omega)}\tau_1 +\epsilon_{d}\tau_3\Big] ,
\label{freegfctscimp}
\end{equation}
where the determinant, $D(\omega):=\det(\underline {G}^0_d(\omega)^{-1})$ is
given by 
\begin{equation}
 D(\omega)=
\omega^2
\Big[1+\frac{\Gamma}{E(\omega)}\Big]^2
-\frac{\Gamma^2\Delta_{\rm sc}^2}{E(\omega)^2}
-\epsilon_{d}^2.
\label{scdet}
\end{equation}
The full Green's function matrix $\underline {G}_d(\omega)^{-1}$ at the
impurity site is given by the Dyson matrix equation 
\begin{equation}
\underline {G}_d(\omega)^{-1}= \underline  G_0^{-1}(\omega)- \underline
\Sigma(\omega),
\label{scgreenfct}
\end{equation}
where we have introduced the self-energy matrix $\underline \Sigma(\omega)$.

\section{Self-energy using the higher $F$-Green's function}
As described by Bulla et al. \cite{BHP98} there is a method to calculate the
self-energy employing a higher $F$-Green's function, and it can also be used for the
case with superconducting bath. In order to derive the equations of motions for the
correlation functions, the identity
\begin{equation}
  \omega\gfbraket{A;B}_{\omega}+\gfbraket{[H,A],B}_{\omega}=\expval{[A,B]_{\eta}}{}
\end{equation}
($\eta=+$ for fermions) is useful.
The calculation taking into account all offdiagonal terms yields
the following matrix equation 
\begin{equation}
\underline  G_0^{-1}(\omega) \underline {G}_d(\omega)-U\underline F(\omega)=\unitop_2,
\label{eomoff}
\end{equation}
with the matrix of higher Green's functions $\underline F(\omega)$,
\begin{equation}
  \underline F(\omega)=
\left(\begin{array}{c c}
F_{11}(\omega) & F_{12}(\omega) \\
F_{21}(\omega) & F_{22}(\omega)
\end{array}\right).
\end{equation}
We have introduced the matrix elements
$F_{11}(\omega)=\gfbraket{\elann{d}{\uparrow}n_{\downarrow};\elcre{d}{\uparrow}}_{\omega}$,
$F_{12}(\omega)=\gfbraket{\elann{d}{\uparrow}n_{\downarrow};\elann{d}{\downarrow}}_{\omega}$, 
$F_{21}(\omega)=-\gfbraket{\elcre{d}{\downarrow}n_{\uparrow};\elcre{d}{\uparrow}}_{\omega}$ and
$F_{22}(\omega)=-\gfbraket{\elcre{d}{\downarrow}n_{\uparrow};\elann{d}{\downarrow}}_{\omega}$.
In the NRG we calculate $F_{11}$ and $F_{21}$ and the others follow from
$F_{12}(\omega)=-F_{21}(-\omega)^*$ and $F_{22}(\omega)=F_{11}(-\omega)^*$.
We can define the self-energy matrix by
\begin{equation}
  \underline \Sigma(\omega)= U \underline F(\omega)\underline
  {G}_d(\omega)^{-1}.
\label{SigF}
\end{equation}
The properties of the Green's function and the higher $F$-Green's function
lead to the relations 
$\Sigma_{12}(\omega)=\Sigma_{21}(-\omega)^*$ and
$\Sigma_{22}(\omega)=-\Sigma_{11}(-\omega)^*$ for the self-energies.
We can therefore calculate the diagonal self-energy 
$\Sigma(\omega)=\Sigma_{11}(\omega)$ and the offdiagonal self-energy
$\Sigma^{\rm off}(\omega)=\Sigma_{21}(\omega)$ and deduce the other two matrix
elements from them.
With the relation (\ref{SigF}) between $\underline G$, $\underline F$ and
$\underline \Sigma$ the Dyson equation (\ref{scgreenfct}) is recovered from (\ref{eomoff}).
Therefore, the Green's function can be calculated from the free Green's
function as given in (\ref{freegfctscimp}) and the self-energy as calculated
from (\ref{SigF}). This scheme will be useful for applications of
dynamical mean field theory with superconducting symmetry breaking,
where the self-energy matrix has to be calculated accurately to find a
self-consistent solution.


\end{appendix}
\section*{References}
\bibliography{artikel,biblio1}
\bibliographystyle{prsty}

\end{document}